\documentclass[12pt]{article}
\pdfoutput=1
\usepackage[a4paper]{geometry}
\usepackage{jheppub, amsmath,amssymb,amsfonts,amsxtra, mathrsfs, makeidx,graphics,graphicx,amsthm,epsfig, bm,longtable,float, color,tikz,mathtools,xfrac,footnote,rotating, lscape, makecell, environ,mathtools, empheq}

\usepackage{subfig}
\usepackage{multirow}
\usepackage{adjustbox}

\usepackage{amsthm}
\usepackage{hyperref}

\usepackage{amsmath}
\usepackage{amsfonts}
\usepackage{commath}
\usepackage[english]{babel}
\usepackage{setspace}

\usepackage{epsfig,amssymb} 
\usepackage{amsmath}

\usepackage{amscd,color}
\usepackage{amsmath,graphicx}
\usepackage{verbatim,booktabs}

\usepackage{latexsym}
\usepackage{amsmath,amsfonts,amssymb,amsthm}
\usepackage{amsmath,amsthm}

\DeclareMathOperator{\Tr}{Tr}
\DeclareMathOperator{\rk}{rk}
\DeclareMathOperator{\sgn}{sgn}
\renewcommand{\Re}{\text{Re}}
\renewcommand{\Im}{\text{Im}}

\title{
\begin{center}
The SCI of $\mathcal{N}=4$ $USp(2N_c)$ and $SO(N_c)$ SYM as a matrix integral
\end{center}}

\author[a]{Antonio Amariti,}
\author[b,c]{Marco Fazzi,} 
\author[a,d]{and Alessia Segati}
\affiliation[a]{INFN, Sezione di Milano, Via Celoria 16, I-20133 Milano, Italy}
\affiliation[b]{INFN, Sezione di Milano-Bicocca, Piazza della Scienza 3, I-20126 Milano, Italy}
\affiliation[c]{Dipartimento di Fisica, Universit\`a di Milano-Bicocca, Piazza della Scienza 3, I-20126 Milano, Italy}
\affiliation[d]{Dipartimento di Fisica, Universit\`a degli Studi di Milano,
Via Celoria 16, I-20133 Milano, Italy}

\emailAdd{antonio.amariti@mi.infn.it,  marco.fazzi@mib.infn.it, alessia.segati@mi.infn.it}

\abstract{We study the superconformal index of 4d $\mathcal{N}=4$ $USp(2N_c)$ and $SO(N_c)$ SYM from a matrix model perspective. We focus on the Cardy-like limit of the index.  Both in the symplectic and orthogonal case the index is dominated by a saddle point solution which we identify,  reducing the calculation to a matrix integral of a pure Chern--Simons theory on the three-sphere.  We further compute the subleading logarithmic corrections, which are of the order of the center of the gauge group. In the $USp(2N_c)$ case we also study other subleading saddles of the matrix integral.  Finally we discuss the case of the Leigh--Strassler fixed point with $SU(N_c)$ gauge group,  and we compute the entropy of the dual black hole from the Legendre transform of the entropy function.}

\begin{document}

\maketitle

%%%%%%%%%%%%%%%%%%%%%%
\section{Introduction}
\label{sec:intro}
%%%%%%%%%%%%%%%%%%%%%%

The 4d superconformal index (SCI), originally defined in \cite{Romelsberger:2005eg,Kinney:2005ej}, is a generalization of the Witten index obtained by radially quantizing a superconformal field theory (SCFT).
It counts a set of protected short multiplets that do not recombine into long ones.
The index can equivalently be obtained by localization on $S^3 \times S^1$ (though the two definitions differ by an overall contribution dubbed supersymmetric Casimir energy in
\cite{Assel:2014paa,Assel:2015nca}.)
The index has been an excellent tool for the study of 4d SCFTs, 
because it is a topological invariant, fully quantum, and protected quantity.
For instance, it has been used to check dualities, propose new ones, study (super)symmetry
enhancements, and analyze the conformal manifold. (See \cite{Rastelli:2016tbz,Gadde:2020yah} and references therein for a recent account on the subject.)

The original motivation behind the introduction of the SCI of $\mathcal{N}=4$ $SU(N_c)$ SYM was counting the 1/16-BPS states that should reproduce the entropy of the dual charged and rotating black hole (BH) in AdS$_5 \times S^5$. (See \cite{Zaffaroni:2019dhb} for a comprehensive review on the subject.)
However this expectation has been puzzling for more than a decade, because the large-$N_c$ index
was found to be an order-one quantity instead of order-$N_c^2$ as expected from the holographic dictionary.
This is due to the large cancellation between fermionic and bosonic states counted by the $(-1)^F$ operator in the index.
A solution to this problem was obtained only recently by noticing that allowing for complex fugacities there is an obstruction to such a cancellation and the dual BH entropy can indeed be extracted from the index.
Two main approaches to the computation of the entropy have been developed. The first one requires an opportune Cardy-like limit 
\cite{Choi:2018hmj,Honda:2019cio,ArabiArdehali:2019tdm,Cabo-Bizet:2019osg,Kim:2019yrz},
while the second approach consists of an exact evaluation of the index in terms of 
(a set of solutions to) the so-called Bethe Ansatz equations (BAE) \cite{Benini:2018mlo,Benini:2018ywd}.\footnote{Even if this second approach is in principle exact, only a set of ``basic'' solutions reproduces the BH entropy.} Many generalizations of these results have since then appeared
\cite{Hosseini:2018dob,Amariti:2019mgp,Larsen:2019oll,Cabo-Bizet:2019eaf,Lanir:2019abx,Murthy:2020rbd,Agarwal:2020zwm,Copetti:2020dil,Goldstein:2020yvj,Hosseini:2020mut}.

An interesting direction regards the calculation of subleading effects that correct the index. Such corrections have been studied in large detail in \cite{GonzalezLezcano:2020yeb} for $\mathcal{N}=4$ $SU(N_c)$ SYM and for the generalization to
$\mathcal{N}=1$ gauge theories representing a stack of D3-branes probing a toric Calabi--Yau threefold singularity.
The calculation has been carried out both in the Cardy-like limit, using a saddle point approximation to the matrix integral, and in the BAE approach, finding agreement between the two descriptions at large $N_c$.
It has been observed that the leading saddle contributing to the index for an $\mathcal{N}=1$ $SU(N_c)$ theory is corrected by a $\log N_c$ term (see \cite[Eq. (3.53)]{GonzalezLezcano:2020yeb}), an appealing result that should be recovered in a supergravity calculation. 
The presence of a $\log N_c$ correction is related to the $\mathbb{Z}_{N_c}$ center symmetry of $SU(N_c)$,
as discussed in \cite{ArabiArdehali:2019orz}.

An analogous calculation in $USp(2N_c)/SO(2N_c+1)$ and $SO(2N_c)$ gauge theories should then yield a 
$\log 2$ and $\log 4$ correction respectively.\footnote{See \cite{Benini:2019dyp,Bobev:2020zov} for similar results in 3d, where the center symmetry determines the logarithmic correction.} In fact these are the dimensions of the centers of the universal covering groups $USp(2N_c)$ and $Spin(N_c)$ (2 or 4 for the latter, for odd and even $N_c$ respectively.)  In all models considered in this paper we only have matter fields in the adjoint representation of the gauge group,  and these do not break the center symmetry.  (Moreover only the gauge algebra is captured by the SCI.\footnote{Keeping this in mind, in the rest of the paper we will be referring to the SCI of $SO(N_c)$ instead of that of $Spin(N_c)$.})
As commented in \cite{Amariti:2021ubd}, for models with other matter representations charged under the center the logarithmic correction corresponds to the  order of the character lattice of the gauge algebra
modulo the action of the Weyl symmetry.  An analogous result has been discussed in \cite{Cassani:2021fyv} in terms of a 
spontaneously broken one-form symmetry.

Motivated by this expectation in this paper we study the logarithmic  corrections to the leading saddle contribution 
to the SCI of 4d $\mathcal{N}=4$ SYM with symplectic and orthogonal gauge group.
We find the expected $\log 2$ and $\log 4$ corrections  to the (logarithm of the) SCI.
As already noted in \cite{GonzalezLezcano:2020yeb}, 
we find that expanding the index in the Cardy-like limit 
one recovers a matrix integral that coincides with the three-sphere partition function 
of a 3d pure Chern--Simons (CS) theory. In the cases at hand the CS theories have gauge group $USp(2N_c)_{\pm (N_c+1)}$, $SO(2N_c+1)_{\pm (2N_c-1)}$, and $SO(2N_c)_{\pm 2(N_c-1)}$ (the subscript representing
the CS level) and this integral can be evaluated exactly. The sign choice is related to a constraint (first discussed in \cite{Hosseini:2017mds,Cabo-Bizet:2018ehj}) satisfied by the chemical potentials appearing in the SCI.

Furthermore in the $USp(2N_c)$ case we analyze in more detail the solutions of the saddle point equations
finding other subleading saddles. We analyze the Cardy-like limit for these solutions
as well.

All the models studied in this paper are examples of 4d \emph{nontoric} gauge theories. Another interesting nontoric theory that we focus on is  the Leigh--Strassler (LS) $\mathcal{N}=1^*$ $SU(N_c)$ fixed point \cite{Leigh:1995ep},  for which we extract the contribution of the leading saddle to the index in the Cardy-like limit.
We find that the entropy function, yielding the entropy of the holographic dual BH after a Legendre transform, is consistent with the result expected from the literature 
\cite{Hosseini:2017mds,Choi:2018hmj,Benini:2018ywd,Honda:2019cio,ArabiArdehali:2019tdm,Cabo-Bizet:2019osg,Kim:2019yrz,Amariti:2019mgp,Lanir:2019abx,Lezcano:2019pae,Benini:2020gjh}, i.e.~is formally obtained  
from the 4d central charge $a$. Furthermore, we extract the $\log N_c$ correction,  consistently with the one obtained for the parent $\mathcal{N}=4$ $SU(N_c)$ SYM. 

This paper is structured as follows.  In section \ref{sec:review} we calculate the Cardy-like limit of the $\mathcal{N}=4$ SCI for all classical gauge groups except $SU(N_c)$. In section \ref{sec:USp} we focus on the $USp(2N_c)$ case, computing dominant contribution and subleading correction for the leading (and other subleading) saddle(s). In section \ref{sec:SO2N+1} we focus on the $SO(2N_c+1)$ odd case, while in section \ref{sec:SO2N} on the $SO(2N_c)$ even case. In section \ref{sec:N=1*} we compute the Cardy-like limit of the SCI of the $\mathcal{N}=1^*$ $SU(N_c)$ LS fixed point.  We present our conclusions in section \ref{sec:conc}. Appendix \ref{app:3d} contains technical details on the calculation of 3d pure CS partition functions.

%%%%%%%%%%%%%%%
\section{\texorpdfstring{The Cardy-like limit of the superconformal index of $\mathcal{N}=4$}{The Cardy-like limit of the superconformal index of N=4}}
\label{sec:review}
%%%%%%%%%%%%%%%

In this section we give a brief review of the strategy that we use in the rest of the paper.  The goal is to expand the SCI in the Cardy-like limit in order to extract the dominant contribution and the logarithmic correction, which will turn out to be of the order of the center of the gauge group $G$.

The SCI is defined as the trace
\begin{equation}
\mathcal{I}_\text{sc} \equiv \Tr(-1)^F e^{-\beta  H_{S^3 \times S^1}} p^{J_1+\frac{R}{2}} q^{J_2 + \frac{R}{2}}
\prod_{b=1}^{\rk F} v_b^{Q_b} 
\end{equation}
where $J_i$ are the angular momenta on the three-sphere, $R$ is the R-charge and $q_b$ (not to be confused with $q$) are
the  conserved charges commuting with the supercharges,
where the index $b$ runs over the Cartan subgroup of the flavor symmetry group $F$,
$b=1,\dots, \rk F$. The quantities $p, q$ and $v_b$
are the associated fugacities.
The index of a gauge theory takes the form
\begin{equation}
\label{indexgen}
\mathcal{I}_\text{sc} = \frac{(p;p)_\infty^{\rk G} (q;q)_\infty^{\rk G}}{|\text{Weyl}(G)|}
\oint_{T^{\rk G}}  \prod_{i=1}^{\rk G} \frac{dz_i}{2 \pi i z_i} 
\frac
{\prod_{a=1}^{n_\chi}
\prod_{\rho_a} 
\Gamma_e((pq)^{{R_a}/{2}} z^{\rho_a} v^{\omega _a})
}
{\prod_\alpha \Gamma_e(z^\alpha)}
\end{equation}
where $\rho_a$ ($\omega_a$) runs over the weights of the representation of the gauge (flavor) group of the $a$-th $\mathcal{N}=1$ matter multiplet ($n_\chi$ being their total number), and $\alpha$ runs over the simple roots of the gauge algebra.
The holonomies $z_i$ are defined on the unit circle,  and the index $i$ runs over the Cartan subgroup of the gauge symmetry group $G$, $i=1,\dots, \rk G$. The quantities $(a;b)_\infty$ are $q$-Pochhammer symbols,
$
(a;b)_\infty \equiv \prod_{k=0}^\infty (1-a b^k)
$,
and $\Gamma_e$ are elliptic Gamma functions,
\begin{equation}
\Gamma_e(z;p,q) =\Gamma_e(z) \equiv \prod_{j,k=0}^\infty 
\frac{1-p^{j+1}q^{k+1} /z}{1-p^j q^k z}\ .
\end{equation}
Following the strategy of \cite{GonzalezLezcano:2020yeb} we then rewrite the integral formula in terms of 
modified elliptic Gamma functions $\tilde \Gamma$.
This is done by expressing the holonomies and various fugacities as
\begin{equation}
p = e^{2 \pi i \sigma},\quad
q = e^{2 \pi i \tau},\quad
v_b = e^{2 \pi i \xi_b},\quad
z_i = e^{2 \pi i u_i}
\end{equation}
with $u_i\in (0,1]$ and $0\sim 1$.  The R-symmetry chemical potential is given by the relation
\begin{equation}
v_R = \frac{1}{2}(\tau+\sigma)\ .
\end{equation}
The modified elliptic Gamma functions are then 
\begin{equation}
\tilde \Gamma(u;\tau,\sigma) 
=
\tilde \Gamma(u) 
\equiv
 \Gamma_e(e^{2 \pi i u};e^{2 \pi i \tau},e^{2 \pi i \sigma})\ ,
\end{equation}
such that the index {indexgen}) becomes
\begin{equation}
\mathcal{I}_\text{sc} = \frac{(p;p)_\infty^{\rk G} (q;q)_\infty^{\rk G}}{|\text{Weyl}(G)|}
\int  \prod_{i=1}^{\rk G} du_i
\frac
{ \prod_{a=1}^{n_\chi}
\prod_{\rho_a} 
\tilde \Gamma(\rho_a(\vec u) +\Delta_a )}
{\prod_\alpha \tilde \Gamma(\alpha(\vec u))}
\end{equation}
where $\Delta_a \equiv  \omega_a(\vec \xi\,) + R_a v_R$.
There is one chemical potential $\Delta_a$ for each field in the theory, and they must 
satisfy the relations imposed by global symmetries, i.e. each superpotential 
term is uncharged under the flavor symmetry and it has R-charge two.

In this paper we will be interested in 4d $\mathcal{N}=4$ SYM with gauge group $G$ given by
$USp(2N_c)$, $SO(2N_c+1)$, and $SO(N_c)$.
The SCI expressed in terms of modified elliptic Gamma functions in these cases reads:
\paragraph{{$\bullet \ G=USp(2N_c)$}:}
\begin{align}
\mathcal{I}_\text{sc}^{USp(2N_c)} 
=&\ \frac{(p;p)_\infty^{N_c} (q;q)_\infty^{N_c}}{2^{N_c} N_c!}
 \prod_{a=1}^{3} 
\tilde \Gamma^{N_c}(\Delta_a )
\int \prod_{i=1}^{N_c} du_i
\frac
{ \prod_{a=1}^{3} \prod_{i<j} ^{N_c}
\tilde \Gamma(\pm u_i \pm u_j +\Delta_a )}
{\prod_{i<j}^{N_c}  \tilde \Gamma(\pm u_i \pm u_j )}\cdot
\nonumber \\
& \cdot 
\frac
{ \prod_{a=1}^{3} \prod_{i=1}^{N_c} 
\tilde \Gamma(\pm 2 u_i+\Delta_a )}
{\prod_{i=1}^{N_c} \tilde \Gamma(\pm 2 u_i  )}
\end{align}
where we used the shorthand $f(a\pm b) \equiv f(a+b) f(a-b) $ (and likewise for $f(\pm a\pm b)$).

\paragraph{{$\bullet \ G=SO(2N_c+1)$}:}
\begin{align}
\label{MISOo}
\mathcal{I}_\text{sc}^{SO(2N_c+1)} 
=&\ \frac{(p;p)_\infty^{N_c} (q;q)_\infty^{N_c}}{2^{N_c} N_c!}
 \prod_{a=1}^{3} 
\tilde \Gamma^{N_c}(\Delta_a )
\int  \prod_{i=1}^{N_c} du_i
\frac
{ \prod_{a=1}^{3} \prod_{i<j}^{N_c} 
\tilde \Gamma(\pm u_i \pm u_j +\Delta_a )}
{\prod_{i<j}^{N_c}  \tilde \Gamma(\pm u_i \pm u_j )}\cdot
\nonumber \\
&\cdot
\frac
{ \prod_{a=1}^{3} \prod_{i=1}^{N_c} 
\tilde \Gamma(\pm  u_i+\Delta_a )}
{\prod_{i=1}^{N_c} \tilde \Gamma(\pm  u_i  )}\ .
\end{align}

\paragraph{{$\bullet \ G=SO(2N_c)$}:}
\begin{align}
\label{MISOe}
\mathcal{I}_\text{sc}^{SO(2N_c)} 
= \frac{(p;p)_\infty^{N_c} (q;q)_\infty^{N_c}}{2^{N_c-1} N_c!}
 \prod_{a=1}^{3} 
\tilde \Gamma^{N_c}(\Delta_a )
\int  \prod_{i=1}^{N_c} du_i
\frac
{ \prod_{a=1}^{3} \prod_{i<j}^{N_c} 
\tilde \Gamma(\pm u_i \pm u_j +\Delta_a )}
{\prod_{i<j}^{N_c}  \tilde \Gamma(\pm u_i \pm u_j )}\ .
\end{align}%
Next we expand the index in the Cardy-like limit $|\tau| \rightarrow 0$ (at fixed $\arg \tau \in (0,\pi)$) restricting to the case 
$\tau =\sigma$.
In order to evaluate the index in this limit 
it is convenient to rewrite it as a matrix model by introducing an effective action $S_\text{eff}$ through the definition
\begin{equation}\label{eq:effS}
\mathcal{I}_\text{sc} \equiv \frac{1}{|\text{Weyl}(G)|} \int \prod_{i=1}^{\rk G} du_i\, e^{ S_\text{eff}(\vec u;\tau,\Delta)} \ .
\end{equation}
With this definition, we have:
%\begin{itemize}
\begin{align}
&\bullet \ G=USp(2N_c) \text{\bf :} & S_\text{\text{eff}}^{USp(2N_c)} =& \sum_{i \neq j }  \Big(  \Big( \sum_{a=1}^3 \log \tilde \Gamma(u_{ij}^{(\pm)}  + \Delta_a)\Big) + \log \theta_0 (u_{ij}^{(\pm)};\tau)\Big) + \nonumber \\
& &&+\!\! \sum_{i=1}^{N_c}  \Big(  \Big( \sum_{a=1}^3 \log \tilde \Gamma(\pm 2 u_{i}+ \Delta_a)\Big)+\log \theta_0 (\pm 2u_{i};\tau)\! \Big)+ \nonumber \\
&&&+N_c\sum_{a=1}^3 \log \tilde \Gamma( \Delta_a) + 2 N_c \log (q;q)_{\infty}\ .  \label{effUSP}
\end{align}
\begin{align}
&\bullet \ G=SO(2N_c+1) \text{\bf:} & S_\text{eff}^{SO(2N_c+1)} = &  \sum_{i \neq j} \Big(\! \Big(\! \sum_{a=1}^3 \log \tilde \Gamma(u_{ij}^{(\pm)}   + \Delta_a)\Big)+ \log \theta_0 (u_{ij}^{(\pm)};\tau) \! \Big) +\nonumber \\
&&&+\!\! \sum_{i=1}^{N_c}\! \Big(\!  \Big( \! \sum_{a=1}^3 \log \tilde \Gamma(\pm u_{i} + \Delta_a)\! \Big)\!+\! \log \theta_0 (\pm u_{i};\tau) \Big)+\nonumber \\
&&&+ N_c\sum_{a=1}^3 \log \tilde \Gamma( \Delta_a) + 2 N_c \log (q;q)_{\infty}\ . \label{effSOodd}
\end{align}
\begin{flalign}
&\bullet \ G=SO(2N_c) \text{\bf :} & S_\text{eff}^{SO(2N_c)}  =& \sum_{i \neq j} \!\!  \Big(\! \Big(\! \sum_{a=1}^3 \log \tilde \Gamma(u_{ij}^{(\pm)}  + \Delta_a)\Big)+ \log \theta_0 (u_{ij}^{(\pm)};\tau)\Big) +\nonumber \\
&&&+N_c\sum_{a=1}^3   \log \tilde \Gamma( \Delta_a) + 2 N_c \log (q;q)_{\infty}\ . \label{effSOeven}
\end{flalign}
\begin{comment}
\paragraph{{$\bullet \ G=USp(2N_c)$}:}
\begin{align}
\label{effUSP}
N_c^2 S_\text{\text{eff}}^{USp(2N_c)} =& \sum_{i \neq j} \Big( \Big( \sum_{a=1}^3 \log \tilde \Gamma(u_{ij}^{(\pm)}  + \Delta_a)\Big)
+ \log \theta_0 (u_{ij}^{(\pm)};\tau)\Big)\ + \nonumber \\
&+ \sum_{i=1}^{N_c}\Big(  \Big(\sum_{a=1}^3 \log \tilde \Gamma(\pm 2 u_{i}+ \Delta_a)\Big)
+ \log \theta_0 (\pm 2u_{i};\tau) \Big)\ +
\nonumber \\
&+
N_c\sum_{a=1}^3 \log \tilde \Gamma( \Delta_a) + 2 N_c \log (q;q)_{\infty}\ .
\end{align}
\paragraph{{$\bullet \ G=SO(2N_c+1)$}:}
\begin{align}
N_c^2 S_\text{eff}^{SO(2N_c+1)} =& \sum_{i \neq j} \Big( \Big( \sum_{a=1}^3 \log \tilde \Gamma(u_{ij}^{(\pm)}  + \Delta_a)\Big)
+ \log \theta_0 (u_{ij}^{(\pm)};\tau)\Big) + N_c\sum_{a=1}^3 \log \tilde \Gamma( \Delta_a)\ +
\nonumber \\
&+ \sum_{i=1}^{N_c}\Big(  \Big(\sum_{a=1}^3 \log \tilde \Gamma(\pm u_{i} + \Delta_a)\Big)
+ \log \theta_0 (\pm u_{i};\tau) \Big) + 2 N_c \log (q;q)_{\infty}\ .
%\nonumber \\
%&+ N_c\sum_{a=1}^3\Big(  \log \tilde \Gamma( \Delta_a) \Big)+ 2 N_c \log (q;q)_{\infty}\ .
\end{align}
\paragraph{{$\bullet \ G=SO(2N_c)$}:}
\begin{align}
N_c^2 S_\text{eff}^{SO(2N_c)}
 =& \sum_{i \neq  j} \Big( \Big( \sum_{a=1}^3 \log \tilde \Gamma(u_{ij}^{(\pm)}  + \Delta_a)\Big)
+ \log \theta_0 (u_{ij}^{(\pm)};\tau)\Big)\ +
\nonumber \\
&+
N_c\sum_{a=1}^3   \log \tilde \Gamma( \Delta_a) + 2 N_c \log (q;q)_{\infty}\ .
\end{align}
\end{comment}%
In the above expressions we have defined the shorthands
\begin{align}
&\tilde{\Gamma} (u_{ij}^{(\pm)} +\Delta_a ) \equiv \tilde{\Gamma}(u_i + u_j +\Delta_a)\tilde{\Gamma}(u_i - u_j +\Delta_a)\ ,  \label{eq:Gammashort} \\
& \theta_0(u_{ij}^{(\pm)};\tau) \equiv \theta_0(u_{ij}^{(+)};\tau)\, \theta_0(u_{ij}^{(-)};\tau)\ ,
\end{align}
where the  elliptic function $\theta_0$ is defined as
\begin{equation}
\theta_{0} (u;\tau) \equiv \prod_{k=0}^{\infty} (1-e^{2\pi i (u + k\tau)}) (1-e^{2\pi i (-u+(k+1)\tau)})\ ,
\end{equation}
and it satisfies $\log \theta_0(u;\tau) = -  \log \tilde{\Gamma}(u)$.\footnote{To prove this identity, one can follow the steps explained below \cite[Eq. (3.2)]{GonzalezLezcano:2020yeb}.} 

We now define the $\tau$-modded value of a complex $\mathbb{C} \ni u \equiv \tilde u + \tau \check u$ (with $\tilde u,\check u \in \mathbb{R}$):
\begin{equation}
\{u\}_\tau \equiv u -  \lfloor \Re(u)-\cot (\arg \tau)\, \Im(u) \rfloor\ ,
\end{equation}
where $ \lfloor u \rfloor$ is the floor function. For a real number it reduces to the usual modded value $\{\tilde u\} \equiv \tilde u-\lfloor \tilde u \rfloor$, and it satisfies
\begin{equation}
\{u\}_\tau = \{\tilde u\}_\tau + \tau \check{u}\ , \quad \{-u\}_\tau = \begin{cases}1- \{u\}_\tau & \tilde{u} \notin \mathbb{Z}\\-\{u\}_\tau & \tilde{u} \in \mathbb{Z}\end{cases}\ .
\end{equation}
At small $|\tau|$ and fixed $\arg \tau \in (0,\pi)$ we have the following asymptotic formulae (see e.g. \cite[App. A]{GonzalezLezcano:2020yeb}):
\begin{align}
\log \,(q;q)_{\infty} =& -\frac{i \pi}{12} \left(\tau+\frac{1}{\tau}\right)
-\frac{1}{2} \log(-i \tau) + \mathcal{O}\left(e^{\frac{2 \pi \sin (\arg \tau)}{|\tau|}}\right)\ ;
 \\
\log\theta_0(u;\tau) =&\ \frac{\pi i}{\tau} \{u\}_\tau (1-\{u\}_\tau )+
\pi i  \{u\}_\tau 
-\frac{i \pi}{6 \tau}(1+3\tau +\tau^2)\ +
\nonumber \\
&+
\log\left(
\left(1-e^{-\frac{2 \pi i }{\tau}(1-\{u\}_\tau)}\right)
\left(1-e^{-\frac{2 \pi i }{\tau}(\{u\}_\tau)}\right)
\right)
+\mathcal{O}\left(
e^{\frac{2 \pi \sin (\arg \tau)}{|\tau|}}
\right)\ ;
 \\
\log\tilde \Gamma(u) 
=&\ 2 \pi i Q(\{u\}_\tau;\tau)+\mathcal{O} \left( |\tau|^{-1} e^{\frac{2 \pi \sin (\arg \tau)}{|\tau|}
\min (\{\tilde u\},1-\{\tilde u\})}  \right)\ ,
\end{align}
provided $\tilde u \nrightarrow \mathbb{Z}$. We defined the quantity
\begin{equation}
Q(u;\tau) \equiv -\frac{B_3(u)}{6 \tau^2} +\frac{B_2(u)}{2\tau} -\frac{5}{12}B_{1}(\tau) +\frac{\tau}{12}
\end{equation}
in terms of the Bernoulli polynomials
\begin{equation}
B_3(u)=u^3-\frac{3 }{2}u^2+\frac{1}{2}u\ , \quad B_2(u)=u^2-u+\frac{1}{6}\ , \quad B_1(u)=u-\frac{1}{2}\ .
\end{equation}
These expressions allow us to expand the effective actions in $\tau$ for small $|\tau|$, and compute the associated saddle point equations \emph{at leading order}:
\begin{equation}
\label{saddleeqns}
0=\frac{\partial S_\text{eff}(\vec u;\tau,\Delta)}{\partial u_i} =  -\frac{i \pi}{\tau^2} \sum_{a=1}^{n_\chi} \frac{\partial \rho_a (\vec u)}{\partial u_{i}}  B_{2}(\{ \rho_a(\vec u) + \Delta_a \}_\tau)\ .
\end{equation}
This is a set of $\rk_G$ equations, and we look for solutions $\vec{u}$,  namely the saddle points of the matrix model, which contain a constant part and a linear term in $\tau$, i.e.  we make an Ansatz for the solutions of the form
\begin{equation}
\label{solns}
u_{i} = {u_*}_{i} + \bar u_{i} \equiv  {u_*}_{i} + v_{i} \tau\ .
\end{equation}
We do this to capture the terms at finite order in $\tau$ in the expansion.  In fact when we plug this Ansatz back into \eqref{eq:effS}, we obtain leading and subleading contributions in $\tau$, logarithmic corrections as well as finite terms. 
%
%
%
%
%
%%%%%%%%%%%% %%%%%%%%%%
\section{Symplectic gauge group}
\label{sec:USp}
%%%%%%%%%%%%%%%%%%%%%%
%
%
%
%
%
Let us start our analysis with the $USp(2N_c)$ case.
%Roots: $\{\pm 2u_i, \pm u_i \pm u_j \}$ with $i\neq j$ and $i=1,\ldots,N_c$. Also $\rk(USp(2N_c)) = N_c$ and $|W|=2^{N_c} N_c!$. In the limit $p=q$, we have $\overline{\kappa}(USp(2N_c)) = \frac{(q;q)_\infty^{2N_c}}{2^{N_c} N_c!}$. Then for $q \to 1$
The effective action in this case is \eqref{effUSP}. 
In this section we study the solutions to the saddle point equations $\frac{\partial}{\partial u_i} S_\text{eff} = 0$ for the $USp(2N_c)$ case. These equations read:\footnote{Given \eqref{eq:Gammashort}, we have:
\begin{equation}
\log \tilde{\Gamma}(u_{ij}^{(\pm)}) = \log \tilde{\Gamma}(u_{ij}^{(+)})+\log \tilde{\Gamma}(u_{ij}^{(-)}) \sim Q(u_{ij}^{(+)};
\tau)+Q(u_{ij}^{(-)};\tau) \sim B_3(u_{ij}^{(+)})+B_3(u_{ij}^{(-)}) + \ldots \ . \nonumber
\end{equation}
Then in the following equation by $B_2(\{ u_{ij}^{(\pm)} +\Delta_a \}_\tau )$ we mean $B_2(\{ u_i + u_j +\Delta_a \}_\tau )+ B_2(\{ u_i - u_j +\Delta_a \}_\tau )$, and so on.}
\begin{align}
\sum_{a=1}^{3} \sum_{j=1}^{N_c} \Bigg( & B_2(\{u_{ij}^{(\pm)} +\Delta_a\}_\tau) -B_2(\{-u_{ij}^{(\pm)}  +\Delta_a\}_\tau)\ + \nonumber \\
&+ B_2(\{2u_{i} +\Delta_a\}_\tau)-B_2(\{-2u_{i} +\Delta_a\}_\tau \Bigg) = 0\ ,
\end{align}
for $i=1,\ldots,N_c$.  We have found  three sets of solutions.\footnote{Observe that we are not claiming that these are the only solutions; other isolated or continuous (sets of) solutions are possible for nongeneric values of $\Delta_a$, compatibly with the constraint $\sum_a \Delta_a=2$. At any rate we will not investigate such sporadic possibilities.}
\begin{enumerate}
\item[$i)$] The first set consists of $L$ holonomies at $u=0$ and the remaining $K\equiv N_c-L$ at 
$u=\frac{1}{2}$. When studying the $\tau$-expansion of the index for these solutions we will distinguish 
two cases. The first one consists of considering either all the holonomies at $0$ or at $\frac{1}{2}$.
We will see that they give the dominating contribution to the superconformal index, capturing the entropy function 
of the dual rotating  black hole under the holographic correspondence. The other saddles correspond to subleading 
effects in this regime and their contributions are paired, i.e. the contribution of the saddle given by 
$L$ holonomies at $0$ and $K$ holonomies at $\frac{1}{2}$ is equivalent to the contribution of
$K$ holonomies at $0$ and $L$ holonomies at $\frac{1}{2}$.
In the case of $N_c$ even there is also a single solution with $L=K$.
\item[$ii)$] The second set of solutions corresponds to placing $L$ holonomies at $u=\frac{1}{4}$ and the remaining $K=N_c-L$ at $u=\frac{3}{4}$. By a symmetry argument we can actually send $u_i \rightarrow -u_i$, and this is equivalent to considering all the holonomies at $u=\frac{1}{4}$.
\item[$iii)$] The last possibility consists of considering $P$ holonomies at $u=0$, $P$ holonomies at $u=\frac{1}{2}$, and the remaining $Q\equiv N_c-2P$ at $u=\frac{1}{4}$.  Observe that if $Q=0$ (which is possible only for even $N_c$) this case is equivalent to the first with $L=K$.
\end{enumerate}
In the following we expand the effective action $S_\text{eff}$ around these saddles.
%
%
%
%
%
%
%%%%%%%%%%%%%%%%%%%%%%%%
\subsection{\texorpdfstring{Leading saddle: $N_c$ coincident holonomies at $u_i=0$ or $u_i=\frac{1}{2}$}{Leading saddle: Nc coincident holonomies at u=0 or u=1/2}}
\label{sub:N0orN1/2}
%%%%%%%%%%%%%%%%%%%%%%%%
%
%
%
%
%
%
%
The Ansatz for the saddle point in this case is
\begin{equation}\label{eq:saddlesUSpNc}
\vec u = \left\{ u_j^{(m)} =\frac{m}{2} + \overline u_j \equiv \frac{m}{2} + v_j \tau\right\} \quad \text{with} \quad  m=0,1\ ,
\end{equation}
i.e. we have two possible sets of saddle point holonomies, consistently with the fact that the center of $USp(2N_c)$ is  $\mathbb{Z}_2$.
Expanding around the saddle point, the effective action becomes
\begin{align}
\label{finusp0}
S_\text{eff} |_{\vec u =\{0\}_{N_c}\, \text{or}\, \{ \frac{1}{2}\}_{N_c}}
=& -
\frac{(2 i \pi  \eta  (N_c+1)) \sum _{i=1}^{N_c} \overline u_i^2}{\tau ^2}
+\sum _{ j\neq k} \log \left(2 \sin \left(\frac{\pi \overline u_{jk}{}^{(\pm )}}{\tau }\right)\right)\ + \nonumber \\
&+2 \sum _{i=1}^{N_c} \log \left(2 \sin \left(\frac{  2\pi \overline u_{i}}{\tau }\right)\right) \ 
%+ \nonumber \\ 
+ \frac{i \pi  (6-5 \eta ) \left(2 N_c^2+N_c\right)}{12}
\ + \nonumber \\
& -i \pi  N_c^2- \frac{i \pi  N_c (2 N_c+1)}{\tau^2}  \prod_{a=1}^{3 }\left(\Delta _a-\frac{\eta +1}{2}\right)
-N_c \log (\tau )\ .
\end{align}
Making the change of variables $- i \sigma_j\equiv v_j \tau $, the SCI becomes
\begin{equation}\label{eq:USpSCI}
\mathcal{I}_\text{sc}^{USp(2N_c)}
=
2 \tau^{N_c} e^{- i \pi \frac{N_c (2N_c+1)}{2}} \mathcal{I}_0^{USp(2N_c)} Z^{USp(2N_c)_{-\eta(N_c+1)}}_{S^3}\ ,
\end{equation}
where the last contribution corresponds to the three-sphere partition function of a 3d $USp(2N_c)$ pure CS theory at level $-\eta(N_c+1)$.\footnote{Where, as usual, the CS contribution to the partition function for the $USp(2N_c)$ case has  an extra factor of 2 w.r.t.~$SU(N_c)$ due to the normalization of the generators \cite{Willett:2011gp}.} We also defined 
\begin{multline}
\label{finusp}
\mathcal{I}_0^{USp(2N_c)} \equiv  \exp \Bigg(-
\frac{i \pi  N_c (2 N_c+1)}{\tau^2} 
\prod_{a=1}^{3 }\left(\Delta _a-\frac{\eta +1}{2}\right)+ \\ +  \frac{1}{12} i \pi  (6-5 \eta ) \left(2 N_c^2+N_c\right)-i \pi  N_c^2
-N_c \log (\tau ) \Bigg)\ .
\end{multline}
We can evaluate $Z^{USp(2N_c)_{-\eta(N_c+1)}}_{S^3}$ exactly; see formula \eqref{eq:USpZeval2}. Adding the latter to \eqref{finusp} we obtain
\begin{equation}
e^{\frac{i \pi  N_c (2 N_c+1)}{2} }\ ,
\end{equation}
that cancels an analogous contribution in \eqref{finusp0}. 
All in all we we are left with 
\begin{equation}
\mathcal{I}_\text{sc}^{USp(2N_c)} = 2  \exp \left(- \frac{ i\pi N_c (2N_c+1)}{\tau^2} \prod_{a=1}^{3} \left(
\{ \Delta_a\}_\tau - \frac{1+\eta}{2}\right)+\mathcal{O} \left(e^{-\frac{1}{|\tau|}}
\right)
+\ldots  \right) 
\end{equation}
where the ellipsis represents the contributions from other saddles we ignored.  In the following we will evaluate the contributions of these saddles,  i.e.  cases \emph{ii)} and \emph{iii)} described at the beginning of this section.

We see the appearance of the expected $\log 2$ correction to $\log \mathcal{I}_\text{sc}^{USp(2N_c)}$, which is due to the degeneracies of the saddles \eqref{eq:saddlesUSpNc} counted by $m$.

%
%
%
%
%
%
%
%%%%%%%%%%%%%%%%%%%%%%%%
\subsection{\texorpdfstring{$L$  holonomies at $u_i=0$ and $L-N_c$  at $u_i=\frac{1}{2}$}{L  holonomies at u=0 and L-Nc at u=1/2}}
\label{sub:L0K1/2}
%%%%%%%%%%%%%%%%%%%%%%%%
%
%
%
%
%
The next saddle point that we discuss corresponds to an Ansatz with $L$ holonomies at $u=0$
and $K \equiv N_c-L$ holonomies at $u=\frac{1}{2}$.  Expanding around this Ansatz we have
\begin{equation}
\vec u = \begin{cases}  \overline v_i \equiv v_i \tau,\  & \quad i=1,\dots,L  \\
 \overline w_r + \frac{1}{2} \equiv  w_r \tau + \frac{1}{2}\ , & \quad r=1,\dots,K
 \end{cases}.
\end{equation}
The effective action in the limit $|\tau| \rightarrow 0$ can be rearranged as
\begin{align}
&S_\text{eff}|_{\vec u=\left\{ \{0\}_L,\{\frac{1}{2}\}_K \right\}} =  \nonumber \\
&-\frac{2 i \pi}{\tau^2} \left( \eta_1(L-K+1) + \eta_2 K \right) \sum_{i=1}^L 
\overline v_i^2
-\frac{2 i \pi}{\tau^2} \left( \eta_1(K-L+1) + \eta_2 L \right) \sum_{r=1}^K 
\overline w_r^2\ + \nonumber  \\
& + \sum_{i<j} \log\Bigg(2\frac{\sin(  \pm\pi \overline v_{ij}^{(\pm)})}{\tau}\Bigg) + \sum_{r<s} \log\left(2\frac{\sin(  \pm\pi\overline w_{rs}^{(\pm)})}{\tau}\right) + 2 \sum_{i=1}^{L} \log\left(2\frac{\sin(  2\pi \overline v_{i})}{\tau}\right)\ + \nonumber 
\end{align}
\begin{align}
& + 2 \sum_{r=1}^{K} \log\left(2\frac{\sin(  2\pi \overline w_{r})}{\tau}\right)- \frac{i \pi (2(L-K)^2+N_c)}{\tau^2}\prod_{a=1}^{3} \left(\{\Delta_a\} -\frac{1+\eta_1}{2}\right)\ +  \nonumber \\
& - \frac{i \pi LK }{\tau^2}\prod_{a=1}^{3} \left(\{2\Delta_a\} -\frac{1+\eta_2}{2}\right)\  +\nonumber \\
&  +i \pi \left(\frac{ i \pi  (6-5 \eta _1) (2 (K-L)^2+N_c)  }{12}  + \frac{  \left(12-5 \eta _2\right) K L}{3} -  N_c^2   \right) -N_c \log \tau \ ,
\end{align}
where we used the relations
\begin{equation}
\sum_{a=1}^{3} \left\{ \Delta_a \right\}_\tau = 2 \tau + \frac{3+\xi_0}{2}\ , \quad  \sum_{a=1}^{3} \left\{ \frac{1}{2} + \Delta_a \right\}_\tau = 2 \tau + \frac{3+\xi_1}{2}
\end{equation}
and $\xi_0= \pm 1$,  $\xi_1 = \pm 1$. We then defined  
$\eta_1 = \xi_0$, while for $\eta_2$ we used the relation
\begin{equation}
 \sum_{a=1}^{3} \left\{ \Delta_a \right\}_\tau+ \left\{ \frac{1}{2} + \Delta_a \right\}_\tau 
=  \sum_{a=1}^{3} \left( \left\{2 \Delta_a \right\}_\tau+\frac{1}{2}\right)
\end{equation}
such that
\begin{eqnarray}
\sum_{a=1}^{3} \left\{2 \Delta_a \right\}_\tau =
4 \tau + \frac{3+\xi_1+\xi_0}{2} \equiv 4 \tau + \frac{3+\eta_2}{2} \ ,
\end{eqnarray}
providing a definition for $\eta_2$.

Again, changing variables as $- i \sigma_j \equiv v_j \tau $ and  $ - i \rho_r \equiv w_r \tau$, there appears a contribution in the index from the three-sphere partition function of a $USp(2L) \times USp(2K)$  pure CS theory.  The two symplectic groups have  CS levels $k_{USp(2L)} = -\eta_1(L-K+1) - \eta_2 K$ and $k_{USp(2K)} = -\eta_1(K-L +1) -\eta_2 L$. These CS integrals can be evaluated using the results presented in appendix \ref{app:3d}, but the result is not 
particularly illuminating and we do not report it here.

%
%
%
%
%
%
%
%%%%%%%%%%%%%%%%%%%%%%%%
\subsection{\texorpdfstring{$N_c$ coincident holonomies at $u_i=\frac{1}{4}$}{Nc coincident holonomies at u=1/4}}
\label{sub:N1/4}
%%%%%%%%%%%%%%%%%%%%%%%%
%
%
%
%
%
%
The Ansatz for the saddle point in this case is
\begin{equation}
\vec u = \left\lbrace u_j = \frac{1}{4} + \overline u_j = \frac{1}{4}+ v_j \tau \right\rbrace \ .
\end{equation}
Plugging this into the effective action and expanding for $|\tau |\rightarrow 0$, the leading contribution becomes
\begin{align}
\label{efffin14}
 S_\text{eff}|_{\vec u =\{\frac{1}{4}\}_{N_c} } = & - \frac{i \pi}{\tau^2} \Bigg( (\xi_0 N_c + \xi_1(N_c+2) ) \sum_{i=1}^{N_c} \overline u_i^2 - (\xi_0-\xi_1)  \Bigg(\sum_{i=1}^{N_c} \overline u_i \Bigg)^2 \, \Bigg)\ + \nonumber
\end{align}
\begin{align}
&+\sum_{i<j} \log\left(2\frac{\sin(\pm \pi \overline u_{ij})}{\tau}\right) + \frac{i \pi N_c }{\tau^2}\prod_{a=1}^{3} \left(\{\Delta_a\} -\frac{1+\eta_1}{2}\right) \ + \nonumber \\
&- \frac{i \pi (N_c^2+N_c) }{4\tau^2}\prod_{a=1}^{3} \left(\{2\Delta_a\} -\frac{1+\eta_2}{2}\right)+ \frac{5 i \pi  N_c }{12}  (\eta _1-(N_c+1) \eta _2)  + \frac{i \pi N_c}{2} - N_c \log \tau \ ,
\end{align}
where $\xi_0$ and $\xi_1$ are defined as in subsection \ref{sub:L0K1/2}.

Once again, upon changing variables as $-i \sigma_j \equiv v_j \tau $ we see the emergence of the contribution of the three-sphere partition function of a $U(N_c)$ vector multiplet. There is also a CS term, where the $SU(N_c)$ and $U(1)$ factors give different contributions. 
Indeed, using the results of \cite[App. A]{Amariti:2020xqm} we can read off the CS terms from \eqref{efffin14}. While the $SU(N_c)$ factor has level  $k_{SU(N_c)} = - N_c \eta_2 +2 (\eta_1-\eta_2)$, the $U(1)$ term has CS level $k_{U(1)} =2 (\eta_1-\eta_2)(N_c+1)$.
Also in this case the evaluation of the CS integrals does not lead to an illuminating expression and we do not report it here.
 %
%
%
%
%
%
%
%%%%%%%%%%%%%%%%%%%%%%%%
\subsection{\texorpdfstring{$P$  holonomies at $u_i=0$, $P$ at $u_i=\frac{1}{2}$, and $N_c-2P$  at $u_i=\frac{1}{4}$}{P  holonomies at u=0,  P at u=1/2, and Nc-2P at u=1/4}}
\label{sub:P0P1/2Q1/4}
%%%%%%%%%%%%%%%%%%%%%%%%
%
%
%
%
%
The last case that we discuss corresponds to the Ansatz with $P$ holonomies at $u=0$,
$P$ holonomies at $u=\frac{1}{2}$, and the remaining $Q \equiv N_c-2P$ at $u=\frac{1}{4}$:
\begin{equation}
\vec u = \begin{cases} 
 \overline v_i \equiv v_i \tau\ , &  i=1,\dots,P  \\
 \overline w_r + \frac{1}{2} \equiv  w_r \tau + \frac{1}{2}\ , & r=1,\dots,P \\
 \overline z_m + \frac{1}{4} \equiv  z_m  \tau + \frac{1}{4}\ ,  & m=1,\dots,Q
 \end{cases}.
\end{equation}
Expanding around this Ansatz, the effective action in the limit $|\tau| \rightarrow 0$ can be rearranged as
\begin{align}
& S_\text{eff}|_{\vec u=\left\{ \{0\}_P,\{\frac{1}{2}\}_P,\{ \frac{1}{4}\} _{Q} \right\}} = \nonumber \\
&-\frac{i \pi}{\tau^2}
\Bigg[\left(2(P+1) \xi_0 + 2P \xi_2 +Q (\xi_1+\xi_3)\right)  \Bigg(\sum_{i=1}^{P} \overline v_i^2 +\sum_{r=1}^{P} \overline w_i^2 \Bigg)\ + \nonumber \\
& +  (Q (\xi_0 + \xi_2 ) +2 \xi_2+2P (\xi_1+\xi_3) ) \sum_{m=1}^{Q} \overline z_m^2 - (\xi_0-\xi_2)\Bigg(\sum_{m=1}^{Q} \overline z_m \Bigg)^2  \Bigg] \ + \nonumber \\
&+
\sum_{i<j} \log\Bigg(2\frac{\sin(\pm \pi \overline v_{ij}^{(\pm)} )}{\tau}\Bigg) + \sum_{r<s} \log \Bigg(2\frac{\sin(  \pm \pi \overline w_{rs}^{(\pm)})}{\tau}\Bigg)\ + \nonumber \\
&+ 2 \sum_{i=1}^{P} \log\left(2\frac{\sin(  2\pi \overline v_{i})}{\tau}\right)
+ 2 \sum_{r=1}^{P} \log\left(2\frac{\sin(  2\pi \overline w_{r})}{\tau}\right) \nonumber \\
&+\sum_{m<n} \log\left(2\frac{\sin(\pm \pi \overline z_{mn})}{\tau}\right)- 
\frac{i \pi (2P-Q) }{\tau^2}\prod_{a=1}^{3} \left(\{\Delta_a\} -\frac{1+\eta_1}{2}\right)\ + \nonumber 
\end{align}
\begin{align}
&-\frac{i \pi ((2P-Q)^2+Q) }{4 \tau^2}\prod_{a=1}^{3} \left(\{2\Delta_a\} -\frac{1+\eta_2}{2}\right) - \frac{ i \pi  PQ }{4\tau^2}\prod_{a=1}^{3} \left(\{4\Delta_a\} -\frac{1+\eta_4}{2}\right)\ + \nonumber \\
&- i \pi  \left(Q^2+4 P^2\right) + \frac{1}{12} i \pi  \left(6-5 \eta _1\right) (2 P-Q) + \frac{1}{12} i \pi  \left(12-5 \eta _2\right) \left((2 P-Q)^2+Q\right)\ + \nonumber \\
&+\frac{1}{3} i \pi  \left(12-5 \eta _4\right) Q P - N_c \log \tau\ ,
\end{align}
where $\xi_{0,1,2,3} = \pm 1$  are defined by the relations
\begin{eqnarray}
\sum_{a=1}^{3} \left\{\frac{J}{4} +\Delta_a \right\}_\tau = 2 \tau + \frac{3+\xi_{J}}{2}\ ,
\quad
J=0,\ldots,3 \ .
\end{eqnarray}
Furthermore we called  $\eta_1\equiv \xi_0$, $\eta_2 \equiv \xi_0+\xi_2$, and  $\eta_4 \equiv \xi_0+\xi_1+\xi_2+\xi_3$.
 
Changing variables as $- i \sigma_j \equiv v_j \tau $, $- i \rho_r \equiv w_r\tau $ and $- i \lambda_m \equiv z_m\tau $
we recognize in the expansion of the index  a contribution from the three-sphere partition function  of a $USp(2P) \times USp(2P) \times U(Q)$ pure CS theory. The two symplectic groups have the same CS level 
$k_{USp(2P)} =-\frac{1}{2}(2(P+1)\xi_0 +2P\xi_2 +Q(\xi_1+\xi_3))$, while the $SU(Q)$ and the $U(1)$ subgroups of $U(Q)$ have different CS levels,  $ -(Q (\xi_0 + \xi_2 ) +2 \xi_2+2P (\xi_1+\xi_3) $ and $-2(\xi_2+Q\xi_2+P(-\xi_0+\xi_1+\xi_2+\xi_3))$ respectively.
Again the evaluation of the CS integrals does not lead to an illuminating expression and we do not report it here.

%
%
%
%
%
%
%%%%%%%%%%%%%%%%%
%%%%%%%%%%%%%%%%%%%%%
\section{Orthogonal gauge group}
\label{sec:SO}
%%%%%%%%%%%%%%%%%%%%%
%%%%%%%%%%%%%%%%%
%
%
%
%
%
In order to study the orthogonal cases for generic rank we first discuss the SCI of $SO(N_c)$ with $N_c=3,\dots,6$. In fact for these values of $N_c$ the index can be extracted by leveraging the accidental isomorphisms of some classical Lie algebras. 
%\begin{itemize}
\paragraph{{$\bullet\ SO(3)$}:}
 In this case, denoting $y$ the holonomy of $USp(2)$ and $u$ the holonomy
of $SO(3)$ we can make the change of variables $u=2y$ and show, by direct inspection, that 
$\mathcal{I}_\text{sc}^{SO(3)} = \mathcal{I}_\text{sc}^{USp(2)}$.
\paragraph{{$\bullet\ SO(4)$}:}
In this case, denoting $y_{1,2}$ the holonomies of $SU(2) \times SU(2)$ and $u_{1,2}$ the holonomies
of $SO(4)$ we can make the change of variables 
\begin{equation}
u_1 = y_1+y_2\ ,  \quad u_2= y_1-y_2\ ,
\end{equation}
 and show that 
$\mathcal{I}_\text{sc}^{SO(4)} =\mathcal{I}_\text{sc}^{SU(2)} \mathcal{I}_\text{sc}^{SU(2)}$, where the right hand side corresponds to the index of 
two decoupled $\mathcal{N}=4$ $SU(2)$ models.
\paragraph{{$\bullet\ SO(5)$}:}
In this case, denoting $y_{1,2}$ the holonomies of $USp(4)$ and $u_{1,2}$ the holonomies
of $SO(5)$ we can make the change of variables 
\begin{equation}
u_1 =y_1+y_2 \ , \quad u_2 =y_1-y_2
\end{equation}
and show that $\mathcal{I}_\text{sc}^{SO(5)} = \mathcal{I}_\text{sc}^{USp(4)}$.
\paragraph{{$\bullet\ SO(6)$}:}
In this case we can consider the holonomies of $SU(4)$ and enforce the $SU$ constraint explicitly on their definition:
\begin{equation}
\pm(x_i - x_j)\ , \quad i<j\ ; \quad \pm(x_i+x_j + 2 x_k)\ , \quad i \neq j \neq k\ ,
\end{equation}
with $i,j=1,2,3$. The holonomies of $SO(6)$, denoted $u_i$ with $i=1,2,3$, can be mapped to the $SU(4)$ ones by the change of variables
\begin{equation}
u_1 = x_2+x_3 \ , \quad u_2 = x_3+x_1\ ,  \quad u_3 = x_1+x_2 \ ,
\end{equation}
thus showing that  $\mathcal{I}_\text{sc}^{SO(6)} = \mathcal{I}_\text{sc}^{SU(4)}$.
%\end{itemize}

For all $SU$ and $USp$ cases (computed in \cite{GonzalezLezcano:2020yeb} and here respectively) we see that the leading 
contribution always has a logarithmic correction compatible with the formula $\log |\text{center}(G)|$, where by $\text{center}(G)$ we obviously mean the center of the gauge group $G$, i.e. $\mathbb{Z}_{N_c}$ and $\mathbb{Z}_2$ respectively.  
As discussed in the introduction this correction is generically smaller if there are fields charged under the center symmetry (which is not the case for SYM).

Motivated by the above discussion, in this section we study the leading contribution to the Cardy-like limit of the SCI for both the 
$SO(2N_c+1)$ and the $SO(2N_c)$ case.
In the $SO(2N_c+1)$ case we find the same result obtained for the leading contribution of the symplectic case, as predicted by S-duality.
Nevertheless the matching is nontrivial because we have a different number of solutions to the saddle point equations. Only after a careful evaluation of the 3d CS partition function we will have a proper matching of the two indices including the finite logarithmic corrections.

%
%
%
%
%
%%%%%%%%%%%%%%%%%%%%%%%%
\subsection{\texorpdfstring{The $SO(2N_c+1)$ case}{The SO(2Nc+1) case}}
\label{sec:SO2N+1}
%%%%%%%%%%%%%%%%%%%%%%%%
%
%
%
%
%
%
%

We start by studying $SO(2N_c+1)$.
In this case the matrix integral is given by formula \eqref{MISOo}.
We can then study the saddle point equations: 
\begin{align}
\sum_{a=1}^{3}
\sum_{j=1}^{N_c}
\bigg( & B_2(\{ u_{ij}^{(\pm)} +\Delta_a\}_\tau) -B_2(\{- u_{ij}^{(\pm)}  +\Delta_a\}_\tau)\  + \nonumber \\
&+ B_2(\{ u_{i} +\Delta_a\}_\tau)-B_2(\{-u_{i} +\Delta_a\}_\tau) \bigg) = 0\ , \label{saddlesoo}
\end{align}
for $i=1,\ldots, N_c$.
 Here we focus only on the solutions that have been studied in \cite{Honda:2019cio} in the Cardy-like limit.
 In this case the leading saddle corresponds to solution at ${u_*}_j =0 $.  We expand the holonomies around this solution as in \eqref{solns}, i.e. $ u_j = 0+ \overline u_j \equiv   v_j \tau$.
%They are given by the Ansatz
%\begin{equation}
%\label{saddleSOo}
%\vec u = \left\{ u_j^{(m)} =\frac{m}{2} + \overline u_j \equiv \frac{m}{2} + v_j \tau\right\} \quad \text{with} \quad  m=0,1\ ,
%\end{equation}
%again consistently with the fact that the center of $SO(2N_c+1)$ is  $\mathbb{Z}_2$.
Expanding  the effective action around this saddle point  we find
\begin{align}
\label{finusodd0}
S_\text{eff}|_{\vec u =\{0\}_{N_c}}= & - 
\frac{( i \pi  \eta  (2N_c-1)) \sum _{i=1}^{N_c} \overline u_i^2}{\tau ^2}
+\sum _{j\neq k} \log \bigg(2 \sin \bigg(\frac{\pi  \overline u_{jk}^{(\pm )}}{\tau }\bigg)\bigg) \ + \nonumber \\
&+\sum _{j=1}^{N_c} \log \bigg(2 \sin \bigg(\frac{ \pm \pi \overline u_{j}{}}{\tau }\bigg)\bigg)-\frac{i \pi  N_c (2 N_c +1)}{\tau^2}  \prod_{a=1}^{3 }\left(\Delta _a-\frac{\eta +1}{2}\right)\ + \nonumber \\
& +\frac{1}{12} i \pi  (6-5 \eta ) \left(2 N_c^2+N_c\right)-i \pi  N_c^2 -N_c \log (\tau )\ .
\end{align}
Upon changing variables as $- i \sigma_j \equiv  v_j \tau$, the SCI becomes
\begin{equation}\label{eq:SOoddSCI}
\mathcal{I}_\text{sc}^{SO(2N_c+1)} =  \tau^{N_c} e^{- i \pi \frac{N_c (2N_c+1)}{2}} \mathcal{I}_0^{SO(2N_c+1)}
Z^{SO(2N_c+1)_{-\eta (2N_c-1)}}_{S^3}
\end{equation}
where the last contribution corresponds to the three-sphere partition function of a 3d $SO(2N_c+1)$ pure CS theory at level $-\eta(2N_c-1)$. We also defined 
\begin{multline}
\label{finusodd}
\mathcal{I}_0^{SO(2N_c+1)}  \equiv \exp \Bigg(-
\frac{i \pi  N_c (2 N_c+1)}{\tau^2}  
\prod_{a=1}^{3 }\left(\Delta _a-\frac{\eta +1}{2}\right)+ \\ +   \frac{1}{12} i \pi  (6-5 \eta ) \left(2 N_c^2+N_c\right)-i \pi  N_c^2 -N_c \log (\tau ) \Bigg)\ .
\end{multline}
We can evaluate $Z^{SO(2N_c+1)_{-\eta (2N_c-1)}}_{S^3}$ exactly as done in formula \eqref{eq:SOoddZeval}. We finally arrive at
\begin{equation}
\mathcal{I}_\text{sc}^{SO(2N_c+1)} = 2 \exp\left(- \frac{i \pi  N_c (2 N_c+1)}{\tau^2} 
\prod_{a=1}^{3 }\left(\Delta _a-\frac{\eta +1}{2}\right) +\mathcal{O}\left(e^{-\frac{1}{|\tau|}}\right)+\ldots \right)\ ,
\end{equation}
where the ellipsis represents the contributions from other saddles ignored here.  

We observe the appearance of the expected $\log 2$ correction to $\log \mathcal{I}_\text{sc}^{SO(2N_c+1)}$,  which is \emph{not} due to the degeneracy of the saddles as in the $USp(2N_c)$ case but rather to the extra factor of $2$ in the evaluation of the partition function for the pure CS theory; see again \eqref{eq:SOoddZeval}.
%
%
%
%
%
%%%%%%%%%%%%%%%%%%%%%%%%
\subsection{\texorpdfstring{The  $SO(2N_c)$ case}{The SO(2Nc) case}}
\label{sec:SO2N}
%%%%%%%%%%%%%%%%%%%%%%%%
%
%
%
%
%
%
%
We now turn to $SO(2N_c)$. In this case the matrix integral is given by formula \eqref{MISOe}.
We can then study the saddle point equations. We have:
\begin{equation}
\label{saddlesoe}
\sum_{a=1}^{3}
\sum_{j=1}^{N_c}
B_2(\{u_{ij}^{(\pm)} +\Delta_a\}_\tau) -B_2(\{-u_{ij}^{(\pm)}  +\Delta_a\}_\tau) =0\ ,
\end{equation}
for $i=1,\ldots,N_c$. Again, we focus only on the solutions that have been studied in \cite{Honda:2019cio} in the Cardy-like limit. They are given by the Ansatz
\begin{equation}
\label{saddleSOe}
\vec u = \left\{ u_j^{(m)} =\frac{m}{2} + \overline u_j \equiv \frac{m}{2} + v_j \tau\right\} \quad \text{with} \quad  m=0,1\ .
\end{equation}
Expanding  the effective action around the saddle point \eqref{saddleSOe} we find
\begin{align}
\label{finusoeven0}
S_\text{eff}|_{\vec u =\{ \frac{m}{2} \}_{N_c}} = &-\ \frac{( 2i \pi  \eta  (N_c-1)) \sum _{i=1}^{N_c} \overline u_i^2}{\tau ^2}
+\sum _{ j\neq k} \log \left(2 \sin \left(\frac{\pi \overline u_{jk}{}^{(\pm )}}{\tau }\right)\right) \ + \nonumber \\
&-\frac{i \pi  N_c (2 N_c-1)}{\tau^2}  \prod_{a=1}^{3 }\left(\Delta _a-\frac{\eta +1}{2}\right)+\frac{1}{12} i \pi  (6-5 \eta ) N_c\left(2 N_c-1\right)\! + \nonumber \\
&-i \pi  N_c(N_c-1) -N_c \log (\tau )\ .
\end{align}
Upon changing  variables as $-i\sigma_j \equiv v_j \tau$,  the SCI becomes
\begin{equation}\label{eq:SOevenSCI}
\mathcal{I}_\text{sc}^{SO(2N_c)} =2  \tau^{N_c} e^{- i \pi \frac{N_c (2N_c-1)}{2}} \mathcal{I}_0^{SO(2N_c)} Z^{SO(2N_c)_{-2\eta (N_c-1)}}_{S^3}\ ,
\end{equation}
where the last contribution corresponds to the three-sphere partition function of a 3d $SO(2N_c)$ pure CS theory at level $-2\eta(N_c-1)$. We also defined 
\begin{multline}
\label{finusoeven}
\mathcal{I}_0^{SO(2N_c)} \equiv \exp \Bigg(- \frac{i \pi  N_c (2 N_c-1)}{\tau^2} \prod_{a=1}^{3 }\left(\Delta _a-\frac{\eta +1}{2}\right) + \\ + \frac{1}{12} i \pi  (6-5 \eta ) N_c\left(2 N_c-1\right)-i \pi  N_c(N_c-1) -N_c \log (\tau )\Bigg)\ .
\end{multline}
Evaluating $Z^{SO(2N_c)_{-2\eta (N_c-1)}}_{S^3}$ exactly as done in formula \eqref{eq:SOevenZeval} and multiplying it by \eqref{finusoeven} we obtain
\begin{equation}
\mathcal{I}_\text{sc}^{SO(2N_c)} = 4 \exp \left( - \frac{i \pi  N_c (2 N_c-1)}{\tau^2}  \prod_{a=1}^{3 }\left(\Delta _a-\frac{\eta +1}{2}\right) +\mathcal{O}\left(e^{-\frac{1}{|\tau|}}\right) +\ldots \right)
\end{equation}
where the ellipsis represents the contributions from other saddles ignored here.

We observe the appearance of the expected $\log 4$ correction to $\log \mathcal{I}_\text{sc}^{SO(2N_c)}$, which is partly due to the degeneracy of the saddles \eqref{saddleSOe} counted by $m$ and partly due to the extra factor of $2$ in the evaluation of the partition function of the pure CS theory. The final result is consistent with the fact that the center is either $\mathbb{Z}_4$ or
$\mathbb{Z}_2 \times \mathbb{Z}_2$ depending on the parity of $N_c$.

%
%
%
%
%
%
%%%%%%%%%%%%%%%%%%%%%%%%%%%%%%%%%%
\section{A nontoric example: the Leigh--Strassler fixed point}
\label{sec:N=1*}
%%%%%%%%%%%%%%%%%%%%%%%%%%%%%%%%%%
%
%
%
%
%

In this section we study the SCI of the so-called $\mathcal{N}=1^*$ theory of \cite{Leigh:1995ep}, i.e. the theory obtained by turning on a complex mass for one of the  $\mathcal{N}=1$ adjoint chirals in $\mathcal{N}=4$ $SU(N_c)$ SYM,  and flowing to the fixed point. We integrate out the massive field $\Phi_3$ after deforming the superpotential of $\mathcal{N}=4$ accordingly:
\begin{equation}
\label{eq:superpot}
\mathcal{W}_{\mathcal{N}=4}^\text{mass} \sim \Tr \Phi_3 [\Phi_1,\Phi_2] + \Tr \Phi_3^2 \quad \longrightarrow \quad \mathcal{W}_{\mathcal{N}=1^*} \sim \Tr [\Phi_1,\Phi_2]^2\ .
\end{equation}
It is interesting to study this case because this $\mathcal{N}=1$ theory is nontoric, and so far such models
have not been discussed in the literature.\footnote{In effect, \cite{GonzalezLezcano:2020yeb} deals only with \emph{toric} $\mathcal{N}=1$ $SU(N_c)$ quivers and gives leading contribution and logarithmic correction of the SCI in the Cardy-like limit; \cite{Kim:2019yrz}  gives only the \emph{leading} contribution for a \emph{general} (i.e. not necessarily toric) $\mathcal{N}=1$ gauge theory with gauge group $G$ in terms of its central charges $a,c$ (and flavor central charges if present).  Partial progress for general theories has also been made in \cite{Cabo-Bizet:2020nkr,Cabo-Bizet:2019osg}.  We are grateful to D.~Cassani for comments on this point.}

\subsection{Superconformal index}

The $N_c-1$ saddle point equations read
\begin{equation}
\sum_{j =1}^{N_c} B_2(\{ u_{ij}+\Delta_a\}_\tau ) - B_2(\{ u_{Nj}+\Delta_a\}_\tau ) 
-
B_2(\{ u_{ij}+\Delta_a\}_\tau ) + B_2(\{ -u_{Nj}+\Delta_a\}_\tau ) =0\ ,
\end{equation}
and they have the same solutions as those discussed in \cite{GonzalezLezcano:2020yeb}, that we report here:
\begin{equation}
\label{solCC}
u_i=  \frac{m}{N_c} + \frac{I-\frac{C-1}{2}}{C} + v_i \tau %u_{I,i-(N_c/C)I}
\quad
\text{with}
\quad
\sum_{i=1}^{N_c} v_i = 0\ ,
\end{equation}
where $I= \lfloor \frac{i-1}{N_c/C} \rfloor$, with $i=1,\dots,N_c$ and $m=0,\dots,\frac{N_c}{C}-1$, with $C$ an integer divisor of $N_c$. This corresponds to the $K$-gon solution of \cite{Cabo-Bizet:2019osg}, and it can be visualized as $C$
sets each containing $N_c/C$ holonomies, uniformly distributed along the unit interval.

\subsubsection{\texorpdfstring{Leading saddle: $C=1$}{Leading saddle: C=1}}

The leading saddle corresponds to the Ansatz with $C=1$.  In the following we discuss this case explicitly.
The fugacities associated with the adjoints $\Phi_1$ and $\Phi_2$ are denoted $\Delta_1$
and $\Delta_2$ respectively, and the superpotential $\mathcal{W}_{\mathcal{N}=1^*}$ in \eqref{eq:superpot} imposes the constraint $\Delta_1+\Delta_2=1$.
It follows that in this case the constraint on the quantities $\{\Delta_a\}_\tau$ is given by 
\begin{equation}\label{eq:LSconstr}
\{\Delta_1\}_\tau + \{\Delta_2\}_\tau =\tau + 1 + \frac{\eta}{2}
\end{equation}
where $\eta=\pm 1$.
By expanding the index at small $|\tau|$ (and fixed $\arg \tau \in (0,1)$) we obtain:
\begin{align}
\label{finLS}
S_\text{eff}^\text{LS} = & \ -\frac{i \pi \eta}{\tau^2} N_c \left(\sum_{i=1}^{N_c} \overline u_i -\frac{1}{N_c} \sum_{j=1}^{N_c}\overline u_j \right)^2
+\sum_{i \neq j} \log \left( 2 \sin \frac{\pi \overline u_{ij}}{\tau}\right) + \nonumber \\ 
&\ - \frac{i \pi  (N_c^2-1)  \left(\Delta_1 - \frac{1+\eta}{2}\right) \left(\Delta_2 - \frac{1+\eta}{2}\right) \left(\Delta_1+\Delta_2 - (1+\eta)\right) }{\tau ^2}\ + \nonumber \\
&\ + \frac{i\pi}{12} (6-5 \eta ) (N_c^2-1)-\frac{i \pi }{2}  (N_c^2-N_c) -(N_c-1) \log \tau \ ,
\end{align}
where we defined $\overline u_i \equiv v_i \tau$.
Once again, upon the change of variables $ i \sigma_j \equiv \frac{m}{N} + v_j \tau$,  we recognize a 3d pure CS partition function. 
By evaluating the latter on the different $N_c$ saddles the final result is
\begin{equation}
\mathcal{I}_\text{sc}^\text{LS} = N_c \, e^{-\frac{\pi i (N_c^2-1)}{\tau^2}
\left(\Delta_1 - \frac{1+\eta}{2}\right)
\left(\Delta_2 - \frac{1+\eta}{2}\right)
\left(\Delta_1+\Delta_2 - (1+\eta)\right)
+\mathcal{O}(e^{-1/|\tau|})\ +\ \ldots}
\end{equation}
where the ellipsis refers to the contribution of other saddles we ignored.  Notice that the index has the functional structure of the 4d central charge $a$,  that in this case is given by
\begin{equation}
a_\text{LS} = \frac{27}{32} \Delta _1 \Delta _2 \left(\Delta _1+\Delta _2\right)\ .
\end{equation} 
Furthermore we observe the appearance of the expected $\log N_c$ correction to $\log \mathcal{I}_\text{sc}^\text{LS}$,  which is inherited from the parent $\mathcal{N}=4$ $SU(N_c)$ SYM.

\subsubsection{\texorpdfstring{Subleading saddles: $C$-center solutions}{Subleading saddles: $C$-center solutions}}
A similar analysis can be carried out for the $C$-center solutions introduced in \cite{ArabiArdehali:2019orz,GonzalezLezcano:2020yeb}. 
Here we redefine \eqref{solCC} as
\begin{equation}
\label{solCC2}
u_i=  \frac{m}{N_c} + \frac{I-\frac{C-1}{2}}{C} + u_{I,i-(N_c/C)I}\ ,
\end{equation}
by  introducing  the quantity $u_{I,i-(N_c/C)I}$.  The action for the $C$-center solution is given by
\begin{align}
S_\text{eff}^{\text{LS},C} = &
\sum_{a=1}^{2} 
\sum_{I,J=0}^{C-1} \sum_{i,j=1}^{N_c/C }
2\pi i Q\left(\left\{\frac{I-J}{C} + \Delta_a\right\}_\tau + u_{I,i}-u_{J,j};\tau \right)\ + \nonumber \\
& +\sum_{I,J=0}^{C} \sum_{i,j=0}^{N_c/C}
\log\left(\theta_0\left(\frac{I-J}{C} +u_{I,i} -u_{J,j};\tau\right)\right) + \nonumber \\
& +2(N_c-1) \log(q;q)_{\infty} \ .
\end{align}
Using the relations
\begin{equation}
\left\{ \frac{J}{C} + \Delta_1 \right\}_\tau+ \left\{ \frac{J}{C} + \Delta_2 \right\}_\tau = \tau +1+\frac{\xi_J}{2}\ , \quad 
 \{C \Delta_1\}_\tau+ \{C \Delta_2\}_\tau = C \tau +1+\frac{\eta_C}{2}\ ,
\end{equation}
with $\xi_0 =\eta_1$ and $\eta_C = \sum_{J=0}^{C-1} \xi_J$,
we can expand the action for $|\tau|\rightarrow 0$ (and fixed $\arg \tau$) obtaining for the leading terms
\begin{align}
&
S_\text{eff}^{\text{LS},C}
=
-\frac{\pi i}{2 \tau^2} \frac{N_c^2}{C^2} \sum_{I,J=0}^{C-1} \xi_{I-J}\left(
\sum_{i=1}^{N_c/C}u_{I,i} -\sum_{j=1}^{N_c/C}u_{J,j} 
\right)^2\ +
\nonumber \\
&
+\sum_{I=0}^{C-1} \left(-\frac{\pi i \eta_c N_c}{C \tau^2} \sum_{i=1}^{N_c/C} \left(u_{I,i}-\frac{C}{N_c} 
\sum_{j=1}^{N_c/C}u_{I,j} \right)^2
+\sum_{i \neq j}^{N_c/C} \log \left((2 \sin \frac{\pi(u_{I,i} - u_{I,j})}{\tau}\right)
\right)\ +	\nonumber \\
&
-\frac{\pi i N_c^2}{C^3 \tau^2} 
\left( \{C\Delta_1\}_\tau - \frac{1+\eta_c}{2} \right) 
\left( \{C\Delta_2\}_\tau - \frac{1+\eta_c}{2} \right) 
\left( \{C\Delta_1\}_\tau + \{C\Delta_2\}_\tau - (1+\eta_c)\right)\!	 +
\nonumber \\
&
+\frac{\pi i }{\tau^2} 
\left( \{\Delta_1\}_\tau - \frac{1+\eta_1}{2} \right) 
\left( \{\Delta_2\}_\tau - \frac{1+\eta_1}{2} \right) 
\left( \{\Delta_1\}_\tau + \{\Delta_2\}_\tau - (1+\eta_1)\right) \ +
\nonumber \\
&
-\frac{5 \pi i \eta_C N_c^2}{12 C}
+
\frac{\pi i N_c}{2}
-\frac{\pi i (6-5\eta_1)}{12}
-(N_c-1)\log\tau\ .
\end{align}
The calculation of the CS integrals is identical to the one performed in \cite{GonzalezLezcano:2020yeb}
for the $C$-center solution of $\mathcal{N}=4$ $SU(N_c)$ SYM.  

The final result is:
\begin{align}
\mathcal{I}_\text{sc}^{\text{LS}, C} = &\
 \frac{N_c}{C} \,e^{
-\frac{\pi i N_c^2}{C^3 \tau^2} 
\left( \{C\Delta_1\}_\tau - \frac{1+\eta_c}{2} \right) 
\left( \{C\Delta_2\}_\tau - \frac{1+\eta_c}{2} \right) 
\left( \{C\Delta_1\}_\tau + \{C\Delta_2\}_\tau - (1+\eta_c)\right) }\ \cdot \nonumber \\
& \cdot e^{
\frac{\pi i }{\tau^2} 
\left( \{\Delta_1\}_\tau - \frac{1+\eta_1}{2} \right) 
\left( \{\Delta_2\}_\tau - \frac{1+\eta_1}{2} \right) 
\left( \{\Delta_1\}_\tau + \{\Delta_2\}_\tau - (1+\eta_1)\right) 
+\frac{5 \pi i (\eta_1-C \eta_C)}{12}}
\cdot
Z^{U(1)}_{S^3}+
\dots\ ,
\end{align}
where $Z^{U(1)}_{S^3} $ denotes the CS partition function  of the abelian  factors as in \cite{GonzalezLezcano:2020yeb}. 

\subsection{Entropy function and dual black hole entropy}

We conclude the analysis of the LS fixed point by studying the the entropy function $S_E$ 
that represents the $\log$ of the number of states and corresponds to the Legendre
transform of the index. In the holographic dictionary the Legendre
transform of $S_E$ gives the entropy of the dual black hole.
The entropy function can be read off of the logarithm of the SCI, and is thus given by
\begin{equation}
S_E = - \kappa \frac{i \pi  \left(\Delta _1-\frac{\eta +1}{2}\right) \left(\Delta _2-\frac{\eta +1}{2}\right) \left(\Delta _1+\Delta _2- (\eta +1)\right)}{\tau ^2}
\end{equation}
with the constraint $\Delta_1+\Delta_2 - \tau -1 - \frac{\eta}{2}=0$ (which is derived from \eqref{eq:LSconstr}).
The overall constant $\kappa$ is fixed as $\kappa=\frac{1}{8}$ (see the discussion in \cite{Benini:2020gjh}).

The entropy is computed in terms of the charges $Q_{1,2}$ and angular momentum $J$ of the dual black hole. (Observe that since we are identifying $\sigma$ and $\tau$,  we only have one angular momentum $J_1=J_2 \equiv J$.)
The Legendre transform of the entropy function $S_E$ is given by the formula
\begin{equation}
S = S_E + 2 \pi i (Q_1 \Delta_1+Q_2 \Delta_2+ J \tau) + 2 \pi i \Lambda \left(\Delta_1+\Delta_2 - \tau -1 - \frac{\eta}{2}\right) \ ,
\end{equation}
where $\Lambda$ is a Lagrange multiplier that enforces the above constraint between the chemical potentials.
The entropy function satisfies the simple equation
\begin{equation}
S_E = \Delta_1 \frac{\partial S_E}{\partial \Delta_1} 
+ \Delta_2 \frac{\partial S_E}{\partial \Delta_2} 
+ \tau \frac{\partial S_E}{\partial \tau} \ ,
\end{equation}
implying that the entropy can be extracted from the Lagrange multiplier $\Lambda$ as $S = -2 \pi i \Lambda $.
In order to find an expression for $\Lambda$ we first write down the equations $\partial_{\Delta_{1,2}} S=0$
and $\partial_{\tau} S=0$.
These three equations allow to express the quantities $\Lambda+Q_{1,2}$ and $\Lambda-J$ in terms of 
$\Delta_{1,2}$ and $\tau$.  They read
\begin{align}
\frac{\Lambda +Q_1}{\kappa }&=-\frac{\left(\Delta _2-\frac{\eta +1}{2}\right) \left(2 \Delta _1+\Delta _2-\frac{3}{2}  (\eta +1)\right)}{2 \tau ^2}\ ,
\nonumber \\
\frac{\Lambda +Q_2}{\kappa }&=-\frac{\left(\Delta _1-\frac{\eta +1}{2}\right) \left(\Delta _1+2 \Delta _2-\frac{3}{2}  (\eta +1)\right)}{2 \tau ^2}\ ,
 \\
\frac{\Lambda -J}{\kappa }&=\frac{\left(\Delta _1-\frac{\eta +1}{2}\right) \left(\Delta _2-\frac{\eta +1}{2}\right) \left(\Delta _1+\Delta _2-\eta -1\right)}{\tau ^3}\ .
\nonumber
\end{align}
Using these relations we can find an identity involving $\Lambda$, $Q_{1,2}$, and $J$. In the case of $\mathcal{N}=4$ SYM this is a cubic equation in $\Lambda$; here instead we found
a fifth-order equation in $\Lambda$, which to the best of our knowledge appears for the first time in such a calculation.  It reads:
\begin{multline}
\frac{1}{2} (\Lambda -J)^2 (\Lambda +2 Q_1-Q_2)  (2 \Lambda +Q_1+Q_2) (\Lambda -Q_1+2 Q_2) \ + \\  +
\frac{27}{32} \kappa  (\Lambda -J)^4-\frac{2 (Q_1-Q_2)^2 (\Lambda +Q_1)^2 (\Lambda +Q_2)^2}{\kappa } =0\ .
\end{multline}
Solving this equation in $\Lambda$ yields the entropy $S$ as a function of the charges as explained above.  In order to obtain a sensible result we should also impose that $\Lambda$ is purely imaginary.
In general a fifth order equation with two  imaginary solutions can be written as
\begin{equation}
\Lambda ^5+ c_2 \Lambda ^4+(c_1+c_3) \Lambda ^3+(c_1 c_2+c_4) \Lambda ^2+c_1 c_3 \Lambda +c_1 c_4=0\ ,
\end{equation}
with solutions $\Lambda = \pm i \sqrt{c_1}$.  The coefficients $c_i$ can be expressed in terms $Q_{1,2}$ and $J$. (This is a reality condition on the entropy, which also imposes a constraint among the  charges.) The BH entropy is then given by the following relation:
\begin{equation}\label{eq:BHentropy}
S = -2 \pi i \Lambda =  2 \pi  \sqrt{c_1} = 2 \pi  \sqrt{\frac{\alpha-\sqrt{\alpha^2+32 \, \kappa  \, \beta}}{16 \kappa }}\ ,
\end{equation}
with
\begin{align}
\alpha\equiv &\ \kappa  J (27 \kappa -8 J )+8  (Q_1+Q_2 )  (3 \kappa  J+4  (Q_1-Q_2 )^2 )+12 \kappa   (Q_1^2-4 Q_2 Q_1+Q_2^2 )\ ,
\nonumber \\
\beta \equiv &\ 27 \kappa ^2 J^3+12 \kappa  J^2 (Q_1^2-4 Q_2 Q_1+Q_2^2)+8 (Q_1+Q_2) (\kappa  J (Q_2-2 Q_1) (Q_1-2 Q_2) \ + \nonumber \\
&+4 Q_1 Q_2 (Q_1-Q_2)^2)\ .
\end{align}

%
%
%
%
%%%%%%%%%%%%%%%%%%%%%%%
\section{Further directions}
\label{sec:conc}
%%%%%%%%%%%%%%%%%%%%%%%
%
%
%
%
%

Here we are going to present some open questions that should be further explored.

First it should be possible to apply the analysis of \cite{ArabiArdehali:2019orz} to classify
the  saddle point solutions of the $USp(2N_c)$ case via its center symmetry,  and 
to relate them to the  massive and Coulomb
vacua of $\mathcal{N}=1^*$ $USp(2N_c)$ SYM on $\mathbb{R}^3 \times S^1$ \cite{Witten:1997bs,Davies:2000nw,Poppitz:2008hr,Anber:2014lba,Bourget:2015lua,Bourget:2016yhy}.
This can be helpful also for the analysis of the  
SCI for both orthogonal and symplectic gauge group from the BAE approach of
\cite{Benini:2018mlo,Benini:2018ywd}. This analysis should provide a useful check of our results.

It is important to complete our analysis for the orthogonal gauge groups, finding other Ans\"atze for the holonomies (in analogy with the $USp(2N_c)$ case) that we did not discuss here, by solving the  saddle point equations \eqref{saddlesoo} and \eqref{saddlesoe}.

Another open question regards the identification of the holographic dual to the finite-order  logarithmic corrections we found. It would be very interesting to obtain this result from the supergravity side. The problem is very similar to the 3d one recently discussed in \cite{Bobev:2020zov}.

We conclude by observing that the \emph{ordinary} large-$N$ limit  of the SCI of the LS
fixed point has been recently studied in \cite{Bobev:2020lsk} from the holographic perspective. 
It would be interesting to construct the dual BH for the LS fixed point in supergravity.
One should be able to reproduce the BH entropy obtained in \eqref{eq:BHentropy} by starting from the truncations discussed in 
\cite{Khavaev:1998fb,Bobev:2014jva}.

%%%%%%%%%%%%%%%%%%
%
%
%
%
%
%
%%%%%%%%%%%%%%%%%%
\section*{Acknowledgments}
%%%%%%%%%%%%%%%%%%
%
%
We wish to thank F.~Benini, A.~Cabo-Bizet, D.~Cassani, and L.~A.~Pando Zayas for useful correspondence and comments on the draft. This work has been supported in part by the Italian Ministero dell'Istruzione, 
Universit\`a e Ricerca (MIUR), in part by Istituto Nazionale di Fisica Nucleare (INFN)
through the “Gauge Theories, Strings, Supergravity” (GSS) research project and in
part by MIUR-PRIN contract 2017CC72MK-003. The work of M.F. is supported
in part by the European Union's Horizon 2020 research and innovation programme
under the Marie Sk\l odowska-Curie grant agreement No.~754496 - FELLINI.

\appendix

%
%
%
%
%%%%%%%%%%%%%%%%%%%%%%%%%%%%%%%
\section{Pure Chern--Simons three-sphere partition function}
\label{app:3d}
%%%%%%%%%%%%%%%%%%%%%%%%%%%%%%%
%
%
%
%
Here we collect some results on the calculation of three-sphere partition functions for the pure CS theories encountered in the main text.

\subsection{$USp(2N_c)$}
\label{appsoo}

Pick two complexes $\omega_1,\omega_2$ in the upper half-plane and define $\omega\equiv \tfrac{1}{2}(\omega_1+\omega_2)$. When localizing on the squashed three-sphere $S^3_b$ with one squashing parameter $b$ (to preserve $\mathcal{N}=2$ supersymmetry in 3d), we set $\omega_1=ib, \omega_2=\tfrac{i}{b}$; therefore $\omega_1\omega_2=-1$ and $\omega = \tfrac{i}{2}(b+\tfrac{1}{b})$.   For the round three-sphere, which we will focus on hereafter, $\omega_1=\omega_2=\omega=i$.  The localization procedure produces hyperbolic Gamma functions
\begin{equation}\label{eq:hypgam}
\Gamma_h(z;\omega_1,\omega_2) \equiv \prod_{m,n=1}^\infty \frac{(n+1)\omega_1+(m+1)\omega_2 - z}{n\omega_1 + m\omega_2}
\end{equation}
from the one-loop determinants for the vector (and matter) multiplets. Then the partition function of pure CS theory (i.e. without matter) with gauge group $USp(2N_c)_k$ and CS level $k=\frac{t}{2}$ is given by a matrix integral $J_{N_c,0,t}$ which has been studied in the mathematical literature. The exact evaluation is given by \cite[Prop. 5.3.18]{VanDeBult}:
\begin{align}\label{eq:USpZ}
J_{N_c,0,t} \equiv &\ \frac{1}{2^{N_c} N_c! } \int \frac{\prod_{j=1}^{N_c} d \sigma_i \, c(2 t \sum_{j=1}^{N_c} \sigma_j^2) }{\prod_{1 \leq i <j \leq N_c} \Gamma_h(\pm \sigma_i \pm \sigma_j) \prod_{j=1}^{N_c} \Gamma_h(\pm \sigma_j)}\\ 
=&\ \frac{e \left(-\frac{(2+\sgn(t)) N_c}{8}\right)}{(t \, \sgn(t))^\frac{N_c}{2}} \,
c \left( -\frac{N_c(N_c+1)(2N_c+1)(\omega_1^2 + \omega_2^2)}{3t}\right) \cdot \nonumber  \\
&\ \cdot \prod_{1 \leq i < j \leq N_c} 4 \sin\left( \frac{\pi(i \pm j)}{t}\right)
\prod_{j=1}^{N_c}2 \sin \left( \frac{2 \pi j}{t}\right)\ , \label{eq:USpZeval}
\end{align}
where
\begin{equation}\label{eq:shiftedexp}
c(z) \equiv e^{\frac{\pi i z}{2\omega_1 \omega_2}} \ , \quad e(z) \equiv e^{2\pi i z}\ .
\end{equation}
From \eqref{eq:USpSCI} we see that in our case $2k=t=2\eta (N_c+1)$ with $\eta = \pm 1$; thus
\begin{equation}
\prod_{1 \leq i < j \leq N_c} 4 \sin\left( \frac{\pi(i \pm j )}{  2 \eta (N_c+1)}\right)
\prod_{j=1}^{N_c}2 \sin \left( \frac{2 \pi j}{ 2 \eta (N_c+1)}\right)
\end{equation}
evaluates to
\begin{equation}
e^{-\frac{i \pi}{2}(N_c^2-N_c)}  (2(N_c+1))^\frac{N_c}{2}(\sgn(\eta))^{N_c^2}\ .
\end{equation}
Notice that this evaluation is nontrivial, so it is interesting to observe that for the nongeneric value of $2k=t$ extracted from the 4d calculation it can be carried out explicitly.  For other values of $t$ one may use the (large-$N_c$) topological string techniques of \cite{Sinha:2000ap} to evaluate $Z^{USp(2N_c)_k}_{S^3}$.

Alternatively, one can work with standard trigonometric functions by exploiting the relation \cite[Eq. (A.18)]{Benini:2011mf}
\begin{equation}
\label{A7}
\frac{1}{\Gamma_h(\pm x)} = -4 \sin\left(\frac{\pi x}{\omega_1} \right) \sin\left(\frac{\pi x}{\omega_2} \right)=- 4 \sinh(\pm \pi x)\ .
\end{equation}
Substituting it back into the integrand of \eqref{eq:USpZ} we gain a factor of $e^{i \pi N_c^2}$.  When the dust settles we are left with:
\begin{align}\label{eq:USpZeval2}
Z^{USp(2N_c)_{\eta(N_c+1)}}_{S^3} %= &\ e^{i \pi N_c^2} J_{N_c,0,2\eta(N_c+1)} \nonumber \\
\equiv &\  \frac{e^{i\pi N_c^2}}{2^{N_c} N_c! } \int \prod_{j=1}^{N_c} d \sigma_j  e^{-2 \eta  \pi i (N_c+1) \sigma_j^2}  (4 \sinh  (\pm \pi \sigma_j))\, \cdot \nonumber \\
  &\cdot   \prod_{1 \leq i <j \leq N_c} 4 \sinh( \pi (\pm \sigma_i \pm \sigma_j)) \nonumber \\
%&=  \frac{1}{2^{N_c} N_c! } \int \prod_{j=1}^{N_c} d \sigma_j  e^{-2 \eta  \pi i (N_c+1) \sigma_j^2}  (2 \sinh \sigma_j)
%\prod_{1 \leq i <j \leq N_c} 2 \sinh \sigma_{ij}^{(\pm)} \nonumber \\
%& 
= &\ e^{i \pi  N_c^2-\frac{5}{12} i \pi  \eta  N_c \left(2 N_c+1\right)} \ .
\end{align}

\subsection{$SO(N_c)$}
\label{appsoe}

Here we evaluate the (round) three-sphere partition function of pure CS theory with gauge group $SO(N_c)_k$.

The partition function $Z^{SO(2N_c+1)_k}_{S_b^3}$ of the $SO(2N_c+1)_k$ pure CS theory on the squashed sphere
is given by the integral
\begin{equation}\label{eq:SOoddZ}
Z^{SO(2N_c+1)_k}_{S_b^3} \equiv
\frac{1}{2^{N_c} N_c!} 
\int \frac{ \prod_{j=1}^{N_c} d\sigma_i  \, e^{\frac{i \pi  k \sigma_j^2}{\omega_1 \omega_2}}}{\prod_{i<j}^{N_c} \Gamma_h(\pm \sigma_i \pm \sigma_j) 
\prod_{i=1}^{N_c}  \Gamma_h (\pm \sigma_i)}\ .
\end{equation}
In the even $Z^{SO(2N_c)_k}_{S_b^3}$ case, the integral reads instead:
\begin{equation}\label{eq:SOevenZ}
Z^{SO(2N_c)_k}_{S_b^3} \equiv
\frac{1}{2^{N_c-1} N_c!} 
\int  \frac{ \prod_{j=1}^{N_c} d\sigma_i \, e^{\frac{i\pi  k \sigma_j^2}{\omega_1 \omega_2}}}{\prod_{i<j}^{N_c} \Gamma_h(\pm \sigma_i \pm \sigma_j)}\ .
\end{equation}
(In both cases, the dependence on $b$ is through $\omega_1,\omega_2$.)

We start our analysis by computing the integral \eqref{eq:SOoddZ} with $k>0$.
The key formula in order to compute such integral is the generalization of the Weyl character formula for the $B_{N_c}$ algebra, that is
\begin{equation}
\det\left\{2 \sin \left(\frac{\pi(2j-1) \sigma_\ell}{z} \right)\right\}_{1\leq j,\ell\leq N_c}=\prod_{1\leq j<\ell\leq N_c} 4 \sin \left( \frac{\pi(\sigma_j\pm \sigma_\ell)
}{z} \right)\prod_{j=1}^{N_c} 2 \sin \left(  \frac{ \pi \sigma_j}{z} \right)\ .
\end{equation}
Upon using the first relation in \eqref{A7}, the integrand of \eqref{eq:SOoddZ} becomes
\begin{equation}
 \frac{\prod_{j=1}^{N_c} e^{\frac{i \pi k \sigma_j^2}{\omega_1 \omega_2}}}{\prod_{1\leq j<\ell \leq N_c}\Gamma_h(\pm \sigma_j \pm \sigma_\ell) \prod_{j=1}^{N_c} \Gamma_h (\pm \sigma_j)}
=
\prod_{a=1}^{2} \det \left \{\frac{\pi (2j-1)\sigma_\ell}{\omega_a} \right \}_{1\leq j,\ell\leq N_c} \!\!\! e^{-\frac{i \pi k \sum_{j=1}^{N_c} \sigma_j^2}{\omega_1 \omega_2}}\ .
\end{equation}
We further simplify the integral using the relation \cite[Eq. (5.3.20)]{VanDeBult}, i.e.
\begin{multline}
\label{usefulrel}
\int
\det \{ f_j(\sigma_\ell) \}_{1\leq j,\ell \leq N_c}
\det \{ g_j(\sigma_\ell) \}_{1\leq j,\ell \leq N_c}
\prod_{\ell=1}^{N_c }h(\sigma_\ell) 
d \sigma_\ell
 = \\
N_c ! \det \left\{ \int  f_j(\sigma)g_\ell(\sigma)h(\sigma)d\sigma \right\}_{1\leq j,\ell \leq N_c}\ .
\end{multline}
The last integral in \eqref{usefulrel} can be explicitly computed:
\begin{multline}
\int
\sin \left( \frac{ \pi (2j-1) \sigma }{\omega_1}\right)
\sin \left( \frac{ \pi (2\ell-1) \sigma }{\omega_2}\right)
e^{\frac{i \pi k \sigma^2}{\omega_1 \omega_2 }} =\\
2 \sqrt{\frac{\omega_1 \omega_2}{i k}}
e^{-\frac{i \pi ( (2j-1)^2\omega_1^2+(2\ell-1)^2 \omega_2^2)}{2 k \omega_1 \omega_2}} \sin \left(\frac{\pi (2j-1)(2\ell-1)}{2k} \right)\ .
\end{multline}
By further using the relation 
\begin{equation}
\det \left\{ 2 \sin\left(\frac{\pi(2j-1)\ell}{2 k}\right) \right\}_{1\leq j<\ell \leq N_c}
=
\prod_{1\leq j<\ell \leq N_c} 4\sin \left(\frac{\pi (j\pm \ell)}{2 k} \right) \prod_{j=1}^{N_c} 2 \sin \left(\frac{\pi j}{2k} \right)
\end{equation}
we can compute \eqref{eq:SOoddZ}, which gives
\begin{align}
Z^{SO(2N_c+1)_{k>0}}_{S_b^3} =&\
\frac{e^{-\frac{ i \pi N_c(4 N_c^2-1) (\omega_1^2+\omega_2^2)}{12k \omega_1 \omega_2} -\frac{3 i \pi  N_c}{4}}}{k^{N_c/2}}\  \cdot \nonumber \\
& \cdot \!\!\! \! \prod_{1\leq j<\ell \leq N_c} 4 \sin \left( \frac{\pi(j+\ell-1)}{k} \right) \sin \left(\frac{\pi(j-\ell)}{k} \right) 
\prod_{j=1}^{N_c} 2 \sin \left(\frac{ \pi (2j-1)}{2k} \right)\, .\label{eq:SOoddZ2}
\end{align}
An analogous computation can be performed in the case of $k<0$,  obtaining
\begin{align}
Z^{SO(2N_c+1)_{k<0}}_{S_b^3} =&\ 
\frac{e^{-\frac{ i \pi N_c(4 N_c^2-1) (\omega_1^2+\omega_2^2)}{12k \omega_1 \omega_2} -\frac{ i \pi  N_c}{4}} }{(-k)^{N_c/2}}\ \cdot
\nonumber \\
& \cdot  \!\!\! \! \prod_{1\leq j<\ell \leq N_c} 4 \sin \left( \frac{\pi(j+\ell-1)}{k} \right) \sin \left(\frac{\pi(j-\ell)}{k} \right)
\prod_{j=1}^{N_c} 2 \sin \left(\frac{ \pi (2j-1)}{2k} \right)\, . \label{eq:SOoddZ2bis}
\end{align}
Fixing $k=- \eta(2N_c-1)$ and $\omega_1 = \omega_2 =i$, and  substituting \eqref{A7} into the integrand of \eqref{eq:SOoddZ}, the partition function evaluates to
\begin{equation}\label{eq:SOoddZeval}
Z^{SO(2N_c+1)_{-\eta(2N_c-1)}}_{S^3} = 2 \,  e^{\frac{5 i \pi  \eta N_c }{12}  (2 N_c+1 ) +i \pi  N_c^2}\ .
\end{equation}

An analogous computation can be performed for $SO(2N_c)_k$. In this case the generalized Weyl
character formula for the $D_{N_c}$ algebra is
\begin{equation}
\frac{1}{2} \det\left\{ 2 \cos\frac{\pi(2j-1) \sigma_\ell}{z} \right\}_{1\leq j,\ell \leq N_c}
=
\prod_{1\leq j< \ell \leq N_c} 4 \sin \left( \frac{\pi(\sigma_j\pm \sigma_\ell)
}{z} \right)\ .
\end{equation}
By following the steps discussed above the computation is straightforward,  and we obtain
\begin{align}
Z^{SO(2N_c)_{k}}_{S_b^3} =&\ 
\frac{e^{-\frac{i \pi N_c}{4} (2-sign(k))}}{|k|^{N_c/2}}
c \left( -\frac{N_c(N_c-1)(2N_c-1) (\omega_1^2+\omega_2^2)}{3k}\right) \cdot
\nonumber \\ 
&\cdot
\prod_{1 \leq j <\ell <N_c} 4 \sin \left( \frac{\pi(j+\ell-2}{|k|} \right) \sin \left( \frac{\pi(j-\ell)}{|k|} \right) \ .
\end{align}
In this case for $k=-2\eta(N_c-1)$, $\omega_1= \omega_2 =i$, and  substituting \eqref{A7}  into the integrand of \eqref{eq:SOevenZ},  we have:
\begin{equation}\label{eq:SOevenZeval}
Z^{SO(2N_c)_{-2 \eta(N_c-1)}}_{S^3} = 2 \,  e^{\frac{5 i \pi  \eta N_c }{12}  \left(2 N_c-1\right)+i \pi (N_c^2-N_c)}\ .
\end{equation}

\paragraph{Note added in v3.}
We conclude this appendix by commenting on a mistake made in a former version of this manuscript.\footnote{We thank the referee for important observations which made us reconsider the derivation of the results presented in this appendix and section \ref{sec:SO}.}

In that version, the evaluation of the $SO(N_c)_k$ partition function was obtained using the limiting case of the duality  obtained in \cite[Sec. 5.3]{Aharony:2013kma}.  (The duality was originally proposed for $O(N_c)_k$ gauge group and tested numerically at the level of partition functions in \cite{Kapustin:2011gh}; further tests were performed in \cite{Hwang:2011qt,Benini:2011mf,Aharony:2011ci}.) There, 3d $\mathcal{N}=2$ $SO(N)_k$ SQCD with $N_f$ flavors is shown to be dual to $SO(N_f-N+|k|+2)_{-k}$ with $N_f$ dual flavors and mesons. If $N_f=0$, $N=2N_c+1$, and $k=2N_c-1$ the dual rank vanishes. Thus the  integral identity between the electric and the magnetic theory yields an exact evaluation of the electric partition function.

A subtlety arises from the fact that the integral identities for the $SO(N_c)_k$ and $O(N_c)_k$ dualities are formally identical,  whereas the partition functions for these two gauge groups differ in the order of the Weyl group, appearing in the constant prefactor of $Z_{S_b^3}$ (see e.g. the definition \eqref{eq:SOoddZ}). The latter divides both sides of the two integral identities (so that it would not play any role); however in the limiting case where the dual rank vanishes the $O(N_c)_k$ partition function becomes $1$ while the $SO(N_c)_k$ partition function evaluates to $2$ (corresponding to the contribution of the ungauged $\mathbb{Z}_2$ symmetry,  interchanging the identity operators $\mathbf{1}$ and  $-\mathbf{1}$). 

This extra factor of 2 was missing in previous versions, and has been obtained here by a direct evaluation of the integral corresponding to the $SO(N_c)_k$ partition function; see \eqref{eq:SOoddZeval} and \eqref{eq:SOevenZeval}.

\bibliographystyle{JHEP}
\bibliography{ref}

\providecommand{\href}[2]{#2}\begingroup\raggedright\begin{thebibliography}{10}

\bibitem{Romelsberger:2005eg}
C.~Romelsberger, \emph{{Counting chiral primaries in N = 1, d=4 superconformal
  field theories}},
  \href{https://doi.org/10.1016/j.nuclphysb.2006.03.037}{\emph{Nucl. Phys. B}
  {\bfseries 747} (2006) 329}
  [\href{https://arxiv.org/abs/hep-th/0510060}{{\ttfamily hep-th/0510060}}].

\bibitem{Kinney:2005ej}
J.~Kinney, J.~M. Maldacena, S.~Minwalla and S.~Raju, \emph{{An Index for 4
  dimensional super conformal theories}},
  \href{https://doi.org/10.1007/s00220-007-0258-7}{\emph{Commun. Math. Phys.}
  {\bfseries 275} (2007) 209}
  [\href{https://arxiv.org/abs/hep-th/0510251}{{\ttfamily hep-th/0510251}}].

\bibitem{Assel:2014paa}
B.~Assel, D.~Cassani and D.~Martelli, \emph{{Localization on Hopf surfaces}},
  \href{https://doi.org/10.1007/JHEP08(2014)123}{\emph{JHEP} {\bfseries 08}
  (2014) 123} [\href{https://arxiv.org/abs/1405.5144}{{\ttfamily 1405.5144}}].

\bibitem{Assel:2015nca}
B.~Assel, D.~Cassani, L.~Di~Pietro, Z.~Komargodski, J.~Lorenzen and
  D.~Martelli, \emph{{The Casimir Energy in Curved Space and its Supersymmetric
  Counterpart}}, \href{https://doi.org/10.1007/JHEP07(2015)043}{\emph{JHEP}
  {\bfseries 07} (2015) 043}
  [\href{https://arxiv.org/abs/1503.05537}{{\ttfamily 1503.05537}}].

\bibitem{Rastelli:2016tbz}
L.~Rastelli and S.~S. Razamat, \emph{{The supersymmetric index in four
  dimensions}}, \href{https://doi.org/10.1088/1751-8121/aa76a6}{\emph{J. Phys.
  A} {\bfseries 50} (2017) 443013}
  [\href{https://arxiv.org/abs/1608.02965}{{\ttfamily 1608.02965}}].

\bibitem{Gadde:2020yah}
A.~Gadde, \emph{{Lectures on the Superconformal Index}},
  \href{https://arxiv.org/abs/2006.13630}{{\ttfamily 2006.13630}}.

\bibitem{Zaffaroni:2019dhb}
A.~Zaffaroni, \emph{{Lectures on AdS Black Holes, Holography and
  Localization}},  2, 2019,
  \href{https://doi.org/10.1007/s41114-020-00027-8}{DOI}
  [\href{https://arxiv.org/abs/1902.07176}{{\ttfamily 1902.07176}}].

\bibitem{Choi:2018hmj}
S.~Choi, J.~Kim, S.~Kim and J.~Nahmgoong, \emph{{Large AdS black holes from
  QFT}},  \href{https://arxiv.org/abs/1810.12067}{{\ttfamily 1810.12067}}.

\bibitem{Honda:2019cio}
M.~Honda, \emph{{Quantum Black Hole Entropy from 4d Supersymmetric Cardy
  formula}}, \href{https://doi.org/10.1103/PhysRevD.100.026008}{\emph{Phys.
  Rev. D} {\bfseries 100} (2019) 026008}
  [\href{https://arxiv.org/abs/1901.08091}{{\ttfamily 1901.08091}}].

\bibitem{ArabiArdehali:2019tdm}
A.~Arabi~Ardehali, \emph{{Cardy-like asymptotics of the 4d $ \mathcal{N}=4 $
  index and AdS$_{5}$ blackholes}},
  \href{https://doi.org/10.1007/JHEP06(2019)134}{\emph{JHEP} {\bfseries 06}
  (2019) 134} [\href{https://arxiv.org/abs/1902.06619}{{\ttfamily
  1902.06619}}].

\bibitem{Cabo-Bizet:2019osg}
A.~Cabo-Bizet, D.~Cassani, D.~Martelli and S.~Murthy, \emph{{The asymptotic
  growth of states of the 4d $ \mathcal{N}=1 $ superconformal index}},
  \href{https://doi.org/10.1007/JHEP08(2019)120}{\emph{JHEP} {\bfseries 08}
  (2019) 120} [\href{https://arxiv.org/abs/1904.05865}{{\ttfamily
  1904.05865}}].

\bibitem{Kim:2019yrz}
J.~Kim, S.~Kim and J.~Song, \emph{{A 4d N=1 Cardy Formula}},
  \href{https://arxiv.org/abs/1904.03455}{{\ttfamily 1904.03455}}.

\bibitem{Benini:2018mlo}
F.~Benini and P.~Milan, \emph{{A Bethe Ansatz type formula for the
  superconformal index}},
  \href{https://doi.org/10.1007/s00220-019-03679-y}{\emph{Commun. Math. Phys.}
  {\bfseries 376} (2020) 1413}
  [\href{https://arxiv.org/abs/1811.04107}{{\ttfamily 1811.04107}}].

\bibitem{Benini:2018ywd}
F.~Benini and P.~Milan, \emph{{Black Holes in 4D $\mathcal{N}$=4
  Super-Yang-Mills Field Theory}},
  \href{https://doi.org/10.1103/PhysRevX.10.021037}{\emph{Phys. Rev. X}
  {\bfseries 10} (2020) 021037}
  [\href{https://arxiv.org/abs/1812.09613}{{\ttfamily 1812.09613}}].

\bibitem{Hosseini:2018dob}
S.~M. Hosseini, K.~Hristov and A.~Zaffaroni, \emph{{A note on the entropy of
  rotating BPS AdS$_7\times S^4$ black holes}},
  \href{https://doi.org/10.1007/JHEP05(2018)121}{\emph{JHEP} {\bfseries 05}
  (2018) 121} [\href{https://arxiv.org/abs/1803.07568}{{\ttfamily
  1803.07568}}].

\bibitem{Amariti:2019mgp}
A.~Amariti, I.~Garozzo and G.~Lo~Monaco, \emph{{Entropy function from toric
  geometry}},  \href{https://arxiv.org/abs/1904.10009}{{\ttfamily 1904.10009}}.

\bibitem{Larsen:2019oll}
F.~Larsen, J.~Nian and Y.~Zeng, \emph{{AdS$_{5}$ black hole entropy near the
  BPS limit}}, \href{https://doi.org/10.1007/JHEP06(2020)001}{\emph{JHEP}
  {\bfseries 06} (2020) 001}
  [\href{https://arxiv.org/abs/1907.02505}{{\ttfamily 1907.02505}}].

\bibitem{Cabo-Bizet:2019eaf}
A.~Cabo-Bizet and S.~Murthy, \emph{{Supersymmetric phases of 4d $ \mathcal{N} $
  = 4 SYM at large $N$}},
  \href{https://doi.org/10.1007/JHEP09(2020)184}{\emph{JHEP} {\bfseries 09}
  (2020) 184} [\href{https://arxiv.org/abs/1909.09597}{{\ttfamily
  1909.09597}}].

\bibitem{Lanir:2019abx}
A.~Lanir, A.~Nedelin and O.~Sela, \emph{{Black hole entropy function for toric
  theories via Bethe Ansatz}},
  \href{https://doi.org/10.1007/JHEP04(2020)091}{\emph{JHEP} {\bfseries 04}
  (2020) 091} [\href{https://arxiv.org/abs/1908.01737}{{\ttfamily
  1908.01737}}].

\bibitem{Murthy:2020rbd}
S.~Murthy, \emph{{The growth of the $\frac{1}{16}$-BPS index in 4d
  $\mathcal{N}=4$ SYM}},  \href{https://arxiv.org/abs/2005.10843}{{\ttfamily
  2005.10843}}.

\bibitem{Agarwal:2020zwm}
P.~Agarwal, S.~Choi, J.~Kim, S.~Kim and J.~Nahmgoong, \emph{{AdS black holes
  and finite N indices}},  \href{https://arxiv.org/abs/2005.11240}{{\ttfamily
  2005.11240}}.

\bibitem{Copetti:2020dil}
C.~Copetti, A.~Grassi, Z.~Komargodski and L.~Tizzano, \emph{{Delayed
  Deconfinement and the Hawking-Page Transition}},
  \href{https://arxiv.org/abs/2008.04950}{{\ttfamily 2008.04950}}.

\bibitem{Goldstein:2020yvj}
K.~Goldstein, V.~Jejjala, Y.~Lei, S.~van Leuven and W.~Li, \emph{{Residues,
  modularity, and the Cardy limit of the 4d $\mathcal{N}=4$ superconformal
  index}},  \href{https://arxiv.org/abs/2011.06605}{{\ttfamily 2011.06605}}.

\bibitem{Hosseini:2020mut}
S.~M. Hosseini and A.~Zaffaroni, \emph{{Universal AdS black holes in theories
  with sixteen supercharges and their microstates}},
  \href{https://arxiv.org/abs/2011.01249}{{\ttfamily 2011.01249}}.

\bibitem{GonzalezLezcano:2020yeb}
A.~Gonz\'alez~Lezcano, J.~Hong, J.~T. Liu and L.~A. Pando~Zayas,
  \emph{{Sub-leading Structures in Superconformal Indices: Subdominant Saddles
  and Logarithmic Contributions}},
  \href{https://arxiv.org/abs/2007.12604}{{\ttfamily 2007.12604}}.

\bibitem{ArabiArdehali:2019orz}
A.~Arabi~Ardehali, J.~Hong and J.~T. Liu, \emph{{Asymptotic growth of the 4d $
  \mathcal{N} $ = 4 index and partially deconfined phases}},
  \href{https://doi.org/10.1007/JHEP07(2020)073}{\emph{JHEP} {\bfseries 07}
  (2020) 073} [\href{https://arxiv.org/abs/1912.04169}{{\ttfamily
  1912.04169}}].

\bibitem{Benini:2019dyp}
F.~Benini, D.~Gang and L.~A. Pando~Zayas, \emph{{Rotating Black Hole Entropy
  from M5 Branes}}, \href{https://doi.org/10.1007/JHEP03(2020)057}{\emph{JHEP}
  {\bfseries 03} (2020) 057}
  [\href{https://arxiv.org/abs/1909.11612}{{\ttfamily 1909.11612}}].

\bibitem{Bobev:2020zov}
N.~Bobev, A.~M. Charles, D.~Gang, K.~Hristov and V.~Reys,
  \emph{{Higher-Derivative Supergravity, Wrapped M5-branes, and Theories of
  Class $\mathcal{R}$}},  \href{https://arxiv.org/abs/2011.05971}{{\ttfamily
  2011.05971}}.

\bibitem{Amariti:2021ubd}
A.~Amariti, M.~Fazzi and A.~Segati, \emph{{Expanding on the Cardy-like limit of
  the superconformal index of the SCI of 4d $\mathcal{N}=1$ ABCD SCFTs}},
  \href{https://arxiv.org/abs/2103.15853}{{\ttfamily 2103.15853}}.

\bibitem{Cassani:2021fyv}
D.~Cassani and Z.~Komargodski, \emph{{EFT and the SUSY Index on the 2nd
  Sheet}},  \href{https://arxiv.org/abs/2104.01464}{{\ttfamily 2104.01464}}.

\bibitem{Hosseini:2017mds}
S.~M. Hosseini, K.~Hristov and A.~Zaffaroni, \emph{{An extremization principle
  for the entropy of rotating BPS black holes in AdS$_{5}$}},
  \href{https://doi.org/10.1007/JHEP07(2017)106}{\emph{JHEP} {\bfseries 07}
  (2017) 106} [\href{https://arxiv.org/abs/1705.05383}{{\ttfamily
  1705.05383}}].

\bibitem{Cabo-Bizet:2018ehj}
A.~Cabo-Bizet, D.~Cassani, D.~Martelli and S.~Murthy, \emph{{Microscopic origin
  of the Bekenstein-Hawking entropy of supersymmetric AdS$_{5}$ black holes}},
  \href{https://doi.org/10.1007/JHEP10(2019)062}{\emph{JHEP} {\bfseries 10}
  (2019) 062} [\href{https://arxiv.org/abs/1810.11442}{{\ttfamily
  1810.11442}}].

\bibitem{Leigh:1995ep}
R.~G. Leigh and M.~J. Strassler, \emph{{Exactly marginal operators and duality
  in four-dimensional N=1 supersymmetric gauge theory}},
  \href{https://doi.org/10.1016/0550-3213(95)00261-P}{\emph{Nucl. Phys. B}
  {\bfseries 447} (1995) 95}
  [\href{https://arxiv.org/abs/hep-th/9503121}{{\ttfamily hep-th/9503121}}].

\bibitem{Lezcano:2019pae}
A.~Gonz\'alez~Lezcano and L.~A. Pando~Zayas, \emph{{Microstate counting via
  Bethe Ans\"atze in the 4d $ \mathcal{N} $ = 1 superconformal index}},
  \href{https://doi.org/10.1007/JHEP03(2020)088}{\emph{JHEP} {\bfseries 03}
  (2020) 088} [\href{https://arxiv.org/abs/1907.12841}{{\ttfamily
  1907.12841}}].

\bibitem{Benini:2020gjh}
F.~Benini, E.~Colombo, S.~Soltani, A.~Zaffaroni and Z.~Zhang,
  \emph{{Superconformal indices at large $N$ and the entropy of AdS$_5$
  $\times$ SE$_5$ black holes}},
  \href{https://doi.org/10.1088/1361-6382/abb39b}{\emph{Class. Quant. Grav.}
  {\bfseries 37} (2020) 215021}
  [\href{https://arxiv.org/abs/2005.12308}{{\ttfamily 2005.12308}}].

\bibitem{Willett:2011gp}
B.~Willett and I.~Yaakov, \emph{{N=2 Dualities and Z Extremization in Three
  Dimensions}},  \href{https://arxiv.org/abs/1104.0487}{{\ttfamily 1104.0487}}.

\bibitem{Amariti:2020xqm}
A.~Amariti and M.~Fazzi, \emph{{Dualities for three-dimensional $\mathcal{N} =
  2$ $SU(N_c)$ chiral adjoint SQCD}},
  \href{https://doi.org/10.1007/JHEP11(2020)030}{\emph{JHEP} {\bfseries 11}
  (2020) 030} [\href{https://arxiv.org/abs/2007.01323}{{\ttfamily
  2007.01323}}].

\bibitem{Cabo-Bizet:2020nkr}
A.~Cabo-Bizet, D.~Cassani, D.~Martelli and S.~Murthy, \emph{{The large-$N$
  limit of the 4d $ \mathcal{N} $ = 1 superconformal index}},
  \href{https://doi.org/10.1007/JHEP11(2020)150}{\emph{JHEP} {\bfseries 11}
  (2020) 150} [\href{https://arxiv.org/abs/2005.10654}{{\ttfamily
  2005.10654}}].

\bibitem{Witten:1997bs}
E.~Witten, \emph{{Toroidal compactification without vector structure}},
  \href{https://doi.org/10.1088/1126-6708/1998/02/006}{\emph{JHEP} {\bfseries
  02} (1998) 006} [\href{https://arxiv.org/abs/hep-th/9712028}{{\ttfamily
  hep-th/9712028}}].

\bibitem{Davies:2000nw}
N.~M. Davies, T.~J. Hollowood and V.~V. Khoze, \emph{{Monopoles, affine
  algebras and the gluino condensate}},
  \href{https://doi.org/10.1063/1.1586477}{\emph{J. Math. Phys.} {\bfseries 44}
  (2003) 3640} [\href{https://arxiv.org/abs/hep-th/0006011}{{\ttfamily
  hep-th/0006011}}].

\bibitem{Poppitz:2008hr}
E.~Poppitz and M.~Unsal, \emph{{Index theorem for topological excitations on
  R**3 x S**1 and Chern-Simons theory}},
  \href{https://doi.org/10.1088/1126-6708/2009/03/027}{\emph{JHEP} {\bfseries
  03} (2009) 027} [\href{https://arxiv.org/abs/0812.2085}{{\ttfamily
  0812.2085}}].

\bibitem{Anber:2014lba}
M.~M. Anber, E.~Poppitz and B.~Teeple, \emph{{Deconfinement and continuity
  between thermal and (super) Yang-Mills theory for all gauge groups}},
  \href{https://doi.org/10.1007/JHEP09(2014)040}{\emph{JHEP} {\bfseries 09}
  (2014) 040} [\href{https://arxiv.org/abs/1406.1199}{{\ttfamily 1406.1199}}].

\bibitem{Bourget:2015lua}
A.~Bourget and J.~Troost, \emph{{Counting the Massive Vacua of N=1* Super
  Yang-Mills Theory}},
  \href{https://doi.org/10.1007/JHEP08(2015)106}{\emph{JHEP} {\bfseries 08}
  (2015) 106} [\href{https://arxiv.org/abs/1506.03222}{{\ttfamily
  1506.03222}}].

\bibitem{Bourget:2016yhy}
A.~Bourget and J.~Troost, \emph{{The Arithmetic of Supersymmetric Vacua}},
  \href{https://doi.org/10.1007/JHEP07(2016)036}{\emph{JHEP} {\bfseries 07}
  (2016) 036} [\href{https://arxiv.org/abs/1606.01022}{{\ttfamily
  1606.01022}}].

\bibitem{Bobev:2020lsk}
N.~Bobev, E.~Malek, B.~Robinson, H.~Samtleben and J.~van Muiden,
  \emph{{Kaluza-Klein Spectroscopy for the Leigh-Strassler SCFT}},
  \href{https://arxiv.org/abs/2012.07089}{{\ttfamily 2012.07089}}.

\bibitem{Khavaev:1998fb}
A.~Khavaev, K.~Pilch and N.~P. Warner, \emph{{New vacua of gauged N=8
  supergravity in five-dimensions}},
  \href{https://doi.org/10.1016/S0370-2693(00)00795-4}{\emph{Phys. Lett. B}
  {\bfseries 487} (2000) 14}
  [\href{https://arxiv.org/abs/hep-th/9812035}{{\ttfamily hep-th/9812035}}].

\bibitem{Bobev:2014jva}
N.~Bobev, K.~Pilch and O.~Vasilakis, \emph{{(0, 2) SCFTs from the
  Leigh-Strassler fixed point}},
  \href{https://doi.org/10.1007/JHEP06(2014)094}{\emph{JHEP} {\bfseries 06}
  (2014) 094} [\href{https://arxiv.org/abs/1403.7131}{{\ttfamily 1403.7131}}].

\bibitem{VanDeBult}
F.~van~de Bult, \emph{{Hyperbolic Hypergeometric Functions,
  http://www.its.caltech.edu/~vdbult/Thesis.pdf}}, {\emph{Thesis} (2008) }.

\bibitem{Sinha:2000ap}
S.~Sinha and C.~Vafa, \emph{{SO and Sp Chern-Simons at large N}},
  \href{https://arxiv.org/abs/hep-th/0012136}{{\ttfamily hep-th/0012136}}.

\bibitem{Benini:2011mf}
F.~Benini, C.~Closset and S.~Cremonesi, \emph{{Comments on 3d Seiberg-like
  dualities}}, \href{https://doi.org/10.1007/JHEP10(2011)075}{\emph{JHEP}
  {\bfseries 10} (2011) 075} [\href{https://arxiv.org/abs/1108.5373}{{\ttfamily
  1108.5373}}].

\bibitem{Aharony:2013kma}
O.~Aharony, S.~S. Razamat, N.~Seiberg and B.~Willett, \emph{{3$d$ dualities
  from 4$d$ dualities for orthogonal groups}},
  \href{https://doi.org/10.1007/JHEP08(2013)099}{\emph{JHEP} {\bfseries 08}
  (2013) 099} [\href{https://arxiv.org/abs/1307.0511}{{\ttfamily 1307.0511}}].

\bibitem{Kapustin:2011gh}
A.~Kapustin, \emph{{Seiberg-like duality in three dimensions for orthogonal
  gauge groups}},  \href{https://arxiv.org/abs/1104.0466}{{\ttfamily
  1104.0466}}.

\bibitem{Hwang:2011qt}
C.~Hwang, H.~Kim, K.-J. Park and J.~Park, \emph{{Index computation for 3d
  Chern-Simons matter theory: test of Seiberg-like duality}},
  \href{https://doi.org/10.1007/JHEP09(2011)037}{\emph{JHEP} {\bfseries 09}
  (2011) 037} [\href{https://arxiv.org/abs/1107.4942}{{\ttfamily 1107.4942}}].

\bibitem{Aharony:2011ci}
O.~Aharony and I.~Shamir, \emph{{On $O(N_c)$ $d = 3$ $\mathcal{N} = 2$
  supersymmetric QCD Theories}},
  \href{https://doi.org/10.1007/JHEP12(2011)043}{\emph{JHEP} {\bfseries 12}
  (2011) 043} [\href{https://arxiv.org/abs/1109.5081}{{\ttfamily 1109.5081}}].

\end{thebibliography}\endgroup

\end{document}